\documentclass[aps,prb,twocolumn,showpacs,floatfix,superscriptaddress]{revtex4}
\usepackage{dcolumn}
\usepackage{bm}
\usepackage{amsmath,amssymb,graphicx}
\usepackage{epstopdf}

\usepackage{graphicx}

\begin{document}

\title{Fundamental Optical Processes in Armchair Carbon Nanotubes}

\author{Erik H.~H\'{a}roz}
\affiliation{Department of Electrical and Computer Engineering, Rice University, Houston, Texas 77005, USA}
\affiliation{The Richard E. Smalley Institute for Nanoscale Science and Technology, Rice University, Houston, Texas 77005}

\author{Juan G.~Duque}
\affiliation{Chemistry Division, Physical Chemistry and Applied Spectroscopy (C-PCS), Los Alamos National Laboratory, Los Alamos, New Mexico 87545, USA}

\author{Xiaomin Tu}
\author{Ming Zheng}
\affiliation{Polymer Division, National Institute of Standards and Technology, Gaithersburg, Maryland 20899, USA}

\author{Angela~R.~Hight~Walker}
\affiliation{Optical Technology Division, National Institute of Standards and Technology, Gaithersburg, Maryland 20899, USA}

\author{Robert~H.~Hauge}
\affiliation{The Richard E. Smalley Institute for Nanoscale Science and Technology, Rice University, Houston, Texas 77005}
\affiliation{Department of Chemistry, Rice University, Houston, Texas 77005, USA}

\author{Stephen K.~Doorn}
\affiliation{Center for Integrated Nanotechnologies, Los Alamos National Laboratory, Los Alamos, New Mexico 87545, USA}

\author{Junichiro~Kono}
\email[]{kono@rice.edu}
\thanks{corresponding author.}
\affiliation{Department of Electrical and Computer Engineering, Rice University, Houston, Texas 77005, USA}
\affiliation{The Richard E. Smalley Institute for Nanoscale Science and Technology, Rice University, Houston, Texas 77005}
\affiliation{Department of Physics and Astronomy, Rice University, Houston, Texas 77005, USA}

\date{\today}
\begin{abstract}
Single-wall carbon nanotubes provide ideal model 1-D condensed matter systems in which to address fundamental questions in many-body physics, while, at the same time, they are leading candidates for building blocks in nanoscale optoelectronic circuits. Much attention has been recently paid to their optical properties, arising from 1-D excitons and phonons, which have been revealed via photoluminescence, Raman scattering, and ultrafast optical spectroscopy of semiconducting carbon nanotubes. On the contrary, dynamical properties of metallic nanotubes have been poorly explored, although they are expected to provide a novel setting for the study of electron-hole pairs in the presence of degenerate 1-D electrons.  In particular, ($n$,$n$)-chirality, or ÔarmchairÕ, metallic nanotubes are truly gapless with massless carriers, ideally suited for dynamical studies of Tomonaga-Luttinger liquids.  Unfortunately, progress towards such studies has been slowed by the inherent problem of nanotube synthesis whereby both semiconducting and metallic nanotubes are produced.  Here, we use post-synthesis separation methods based on density gradient ultracentrifugation and DNA-based ion-exchange chromatography to produce aqueous suspensions strongly enriched in armchair nanotubes.  Through resonant Raman spectroscopy of the radial breathing mode phonons, we provide macroscopic and unambiguous evidence that density gradient ultracentrifugation can enrich armchair nanotubes.  Furthermore, using conventional, optical absorption spectroscopy in the near-infrared and visible range, we show that interband absorption in armchair nanotubes is strongly excitonic.   Lastly, by examining the G-band mode in Raman spectra, we determine that observation of the broad, lower frequency (G$^{-}$) feature is a result of resonance with non-armchair ``metallic'' nanotubes.  These findings regarding the fundamental optical absorption and scattering processes in metallic carbon nanotubes lay the foundation for further spectroscopic studies to probe many-body physical phenomena in one dimension.
\end{abstract}

\maketitle


\section{Introduction}

Armchair carbon nanotubes have a distinguished status among all the members, or species, of the single-wall carbon nanotube (SWCNT) family.  Being the only truly gapless species in the family, they are defined by the simple relation $n=m$, where $n$ and $m$ are the chiral indices, ($n,m$), which define the diameter and chiral angle of a given nanotube.  As a result of their unique one-dimensional (1-D) electronic dispersions with zero effective mass and zero band gap, they are expected to exhibit some of the unusual properties characteristic of 1-D metals.\cite{Giamarchi04Book}  While strong electron-electron interactions can lead to the formation of Tomonaga-Luttinger liquid states,\cite{Tomonaga50PTP,Luttinger63JMP,MattisLieb65JMP,Mattis93Book,Voit95RPP,GogolinetAl98Book,Schofield99CP,Mahan00Book,SchulzetAl00Book,Giamarchi04Book} strong electron-phonon interactions can renormalize phonon frequencies and lifetimes (Kohn anomalies) and induce Peierls lattice instabilities, especially in small-diameter nanotubes.  

Although a number of theoretical proposals exist for probing many-body effects in armchair SWCNTs,\cite{KaneetAl97PRL, BalentsEgger00PRL, Balents00PRB, DeMartinoetAl02PRL, LevitovTsvelik03PRL, BelluccietAl05PRL, Guinea05PRL, DoraetAl07PRL, DoraetAl08PRL, MkhitaryanetAl08PRL, MishchenkoStarykh11PRL} experimental realizations have been limited to DC transport measurements.\cite{BockrathetAl99Nature, YaoetAl99Nature, Yaoetal01Book, DeshpandeBockrath08NP, DeshpandeetAl09Science}  Recent theories,\cite{Balents00PRB, MishchenkoStarykh11PRL} in particular, have addressed the issue of Fermi edge singularities, specifically for SWCNTs.  Balents\cite{Balents00PRB} predicts that an orthogonality catastrophe strongly reduces the single-particle van Hove singularity to the form $(\varepsilon - \Delta)^{\beta - 1/2}$, with $\beta \approx$ 0.3 for carbon nanotubes, where $\varepsilon$ is energy and $\Delta$ is the energy gap from the Fermi energy to the first subband, which can be probed via tunneling measurements.  Mishchenko and Starykh\cite{MishchenkoStarykh11PRL} predict that optical absorption between the linear band and the first massive band (the so-called $E^M_{01}$ or $E^M_{10}$ transition) is enhanced, compared to the non-interacting case, and develops a power-law frequency dependence $\propto (\omega - \Delta)^{-\gamma}$, where $\gamma \approx$ 0.2 for typical nanotubes.\cite{DubayetAl02PRL, BohnenetAl04PRL, ConnetableetAl05PRL, BarnettetAl05PRB}  Such predictions provide the theoretical impetus for experimentally probing non-equilibrium many-body responses driven through optical excitations and other quantum effects in armchair SWCNTs.  At the same time, as exceptionally conductive wires, exhibiting ballistic conduction even at room temperature,\cite{TansetAl97Nature, WhiteTodorov98Nature, Yaoetal00PRL} they are promising for a variety of electronic applications.

However, specific and systematic studies of armchair SWCNTs aimed at resolving the aforementioned issues have been severely hampered by sample issues.  Although experimental measurements are typically more facile to implement using macroscopic ensembles, because of the heterogeneity inherent to nanotube synthesis, as-produced, bulk samples normally contain a wide assortment of nanotube species of different diameters, chiral angles, lengths, optical enantiomers and electronic types.  This heterogeneity in structural properties often leads to overlap in electronic and optical properties, resulting in a response that is a superposition of signals from both metallic and semiconducting nanotubes.  Often when such overlap exists, the response from semiconducting SWCNTs can obscure that from metallic nanotubes, leading to only limited and incomplete pictures of optical behavior in these gapless materials as demonstrated by numerous optical studies on ensemble-type samples (e.g., references \cite{NairetAl06AnalChem, HagenHertel03NL}).  The alternative to such macroscopic studies is to perform micro-spectroscopy on isolated, individual SWCNTs.  These studies, however, are typically more elaborate to implement experimentally as the signal tends to be small from a single nanotube.  Additionally, study of a particular species of nanotube, like an armchair SWCNT, is almost dependent on chance as nanotube synthesis cannot be sufficiently controlled to grow specific chiralities.  As a result, researchers are limited to the laborious process of searching many isolated tubes to find one that fits the experimental needs.  Lastly, even in cases where a single example of an armchair SWCNT can be located, constituting a single measurement, the resulting data cannot necessarily be generalized to be representative of all armchair SWCNTs due to possible defects and environmental effects.

Here, we present a route to study armchair SWCNTs on a macroscopic scale using various forms of optical spectroscopy.  Using post-synthesis separation techniques such as density gradient ultracentrifugation (DGU) and DNA-based chromatography, researchers have been able to create macroscopic ensembles strongly enriched in armchair species, which can then be probed optically to reveal the intrinsic optical response of metallic nanotubes.  In Section \ref{bandstructure}, we briefly review the electronic band structure and optical selection rules for armchair carbon nanotubes and the related narrow-gap semiconducting nanotubes which constitute the class of ``metallic" SWCNTs.  We then in Section \ref{previous} review the previous experimental studies done on ``metallic" SWCNTs using optical absorption, resonant Raman scattering, and resonant Rayleigh scattering spectroscopies.  In Section \ref{Separation}, we discuss how to prepare macroscopic ensembles using post-growth separation techniques.  DGU (Section \ref{DGU}) provides colored aqueous suspensions consisting of ensembles of multiple armchair species.  The DNA-based method (Section \ref{DNA}) goes further to produce suspensions of a single-chirality armchair species.  Such highly enriched and specialized samples are perfect playgrounds for studying the optical properties of armchair SWCNTs.  Through optical absorption (Section \ref{absorption}), we assess the optical structure including line shape, line width and line position of optical transitions in armchair SWCNTs.  Based on analysis of both DGU and DNA-based armchair samples, we comment on the importance of excitons to the optical properties of metallic SWCNTs and how the role of excitons explains the rainbow of visible colors exhibited by armchair species of different diameters.  Through resonant Raman spectroscopy, we used the excitation profiles of the radial breathing modes (RBMs) to determine optical transition energies and the relative abundances of all ``metallic" nanotube species both before and after DGU enrichment (Section \ref{RRS-RBM}).  While DGU-based separation revealed strong enrichment in armchair species, even ``unsorted" SWCNT material revealed a preference for armchair species.  Extending our studies to higher energy phonons, armchair ensembles revealed the intrinsic line shape of the G-band phonon as originating only from the G$^{+}$ peak, a transverse optical phonon in armchair SWCNTs (Section \ref{RRS-Gband}).  Lastly, in Section \ref{outlook} we summarize our results and discuss future avenues of investigation.

\section{Band Structure and Selection Rules for Optical Transitions in Metallic SWCNTs}
\label{bandstructure}

The so-called ``metallic" SWCNTs are defined as any ($n,m$) species where the quantity ($n-m$) is an integer multiple of 3, or in other words,
\begin{equation}
\nu = (n-m) \bmod 3 = 0,
\label{nu eqn}
\end{equation}
where $n$ and $m$ are the chiral indices.  This leads us to the more accurate classification (as will be shown shortly) of ``metallic" species as $\nu$~=~0 species, which can be further subdivided into the armchair ($n=m$) and the non-armchair ($n \neq m$) species.  The diversity in physical crystal structure between different $\nu$~=~0 species results in a rich mixture of electronic band structure effects, crystal symmetries, and quasiparticle interactions that strongly define and affect the optical properties (and their related processes) of armchair and non-armchair SWCNTs.  Here, we briefly summarize the most important and relevant effects as they pertain to optical spectra of $\nu$~=~0 nanotubes.  

\subsection{Band structure of $\nu = 0$ SWCNTs}

In 1992, with the initial discovery of carbon nanotubes having just occurred by Iijima \textit{et al.},\cite{Iijima91Nature} the electronic band structure of SWCNTs and specifically ``metallic" tubes was calculated by multiple groups\cite{SaitoetAl92APL, MintmireetAl92PRL, HamadaetAl92PRL} following different calculation approaches but yielding similar conclusions.  It was shown that SWCNTs could possess either metallic or semiconducting electronic conduction, depending on the crystal structure of the particular type of tube and specified by the diameter and chiral angle of the nanotube (determined by the chiral indices ($n,m$)).  Specifically, the so-called ($n,n$) armchair species were shown to be metals with no energy band gap and a finite electronic density-of-states at the Fermi energy.\cite{SaitoetAl92APL, MintmireetAl92PRL, HamadaetAl92PRL}  This is a consequence of one of the allowed $k$-states of an armchair nanotube passing through the K (K')-point of the two-dimensional graphene Brillouin zone, where the so-called Dirac cones meet.  The zone-folding approach of Saito \textit {et al.}\cite{SaitoetAl92APL} yielded analytical expressions for calculating the electronic dispersion of the various bands resulting from the quantization of the allowed $k$-states of nanotubes (i.e., cutting lines) due to their one-dimensionality.   Using a $k \cdot p$ formalism near the K-point of graphene, Ajiki and Ando\cite{AjikiAndo93JPSJ, AjikiAndo94Physica} derived an analytical expression for the dispersion of the conduction ($+$) and valence ($-$) bands:
\begin{equation}
\varepsilon_{\pm,q}(k) = \pm {3 a_{c-c} \gamma_{0} \over 2} \sqrt{\kappa_{\varphi}(q)^2 +k^2}.
\label{dispersion eqn}
\end{equation}
Here, $\kappa_{\varphi}(q) = (2/d_{t})(q+\varphi + \nu/3)$ where  $\gamma_{0}$ is the nearest neighbor hopping energy, $a_{c-c}$ (= 1.44~$\AA$) is the carbon-carbon bond length, $d_{t}$ is the nanotube diameter, $q$ is the band index, $\nu$ is the type number as defined earlier (which is 0 for ``metallic" species), and $\varphi = \phi/\phi_{0}$ is the magnetic flux threading a nanotube in units of the magnetic flux quantum, $\phi_{0} = ch/e$.  

\begin{figure}
\centering
\includegraphics[scale=.685]{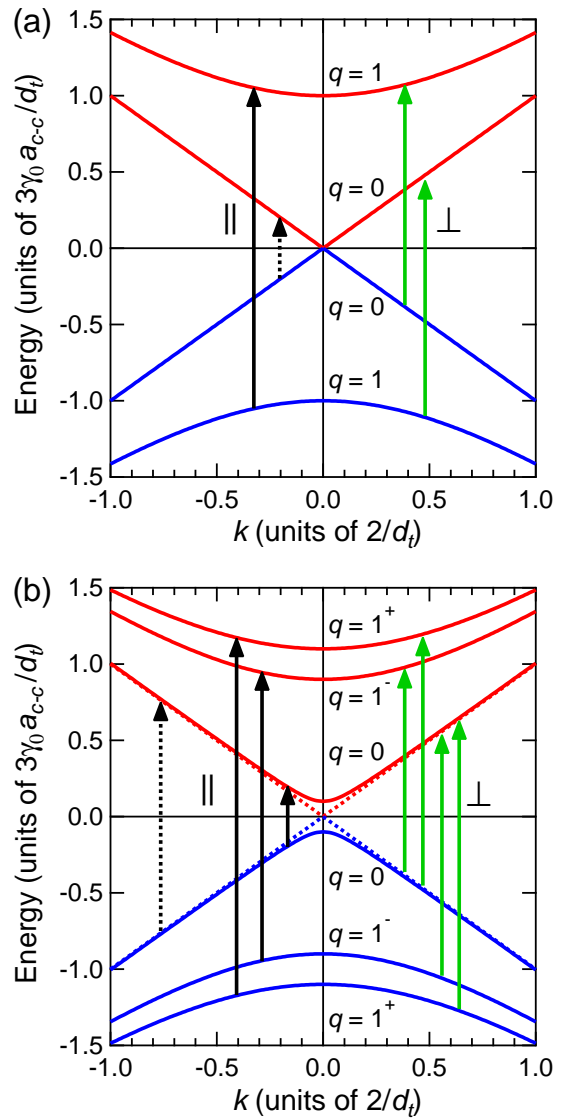}
\caption{Electronic band structure for (a)~armchair or ($n,n$) and (b)~non-armchair, $\nu$~=~0 SWCNTs, where $n \neq m$ [i.e.,~($n,m$)], showing both the massless, linear ($q$~=~0) and massive, hyperbolic ($|q|\geq1$) bands.  Note that in case (b)~for non-armchair $\nu$~=~0 species, the $q=1$ bands split into $q=1^{+}$ and $1^{-}$ bands due to trigonal warping.  Also, the massless $q=0$ bands become massive in the vicinity of $k=0$ due to the curvature-induced band gap.  Valence (conduction) bands are indicated in blue (red).  Solid black (green) arrows indicate allowed optical transitions for electric polarization parallel (perpendicular) to the nanotube axis.  Black dashed arrows indicate forbidden optical transitions.}
\label{fig-MetalBand}
\end{figure}

By considering the effect of nanotube curvature through lattice optimization before calculating band structure, Hamada \textit{et al.} correctly predicted that such non-armchair ``metals" ($\nu$~=~0) were actually narrow-gap semiconductors due to a shift in momentum of the cutting line that would normally pass through the K-point, resulting in the opening of a small band gap.  This shift is a result of enhanced electronic transfer in the circumferential direction of the nanotube due to its curvature.  Kane and Mele\cite{KaneMele97PRL} later derived an analytical expression for this narrow band gap
\begin{equation}
E_{g} = {3\gamma_{0}a_{c-c}^2 \over 4 d_{t}^2} \cos(3\alpha),
\label{curvature eqn}
\end{equation}
where $E_{g}$ is the curvature-induced gap, $\alpha$ is the chiral angle and all other quantities are defined as before.  This curvature-induced band gap is reflected in the electronic band dispersions of ``metallic" species in Fig.~\ref{fig-MetalBand}.  In Fig.~\ref{fig-MetalBand}a, which represents an example of an armchair nanotube, no band gap is observed because the momentum shift is parallel and along the already existing cutting line.  In the case of a non-armchair, $\nu$~=~0 species, a band gap has opened in the linear dispersion bands ($q=0$) due to the curvature effect.

In the vicinity of the K-point of the first Brillouin zone of graphene, all equienergy lines around the K-point are circular around the K-point.  While valid for small-energy excitations, this approximation fails at larger energies when one approaches the M-point where the equienergy lines become triangular.  This distortion of the equienergy contours in the two-dimensional (2D) graphene Brillouin zone from circular to triangular is known as trigonal warping.  As a result, the allowed electronic transitions develop a dependence on chiral angle and electronic type with increasing energy.  This is particularly relevant for $\nu$~=~0 species where initially equidistant cutting lines from the K-point, which intersected the same equienergy line, now become inequivalent due to trigonal warping, resulting in different energy electronic states.  The overall effect is that a single ``metallic" electronic state becomes split into two by a quantity known as the trigonal-warping splitting.\cite{SaitoetAl00PRB}  Using a higher-order $k \cdot p$ calculation,\cite{AjikiAndo96JPSJ, Ando04JPSJ, UryuAndo08PRB} the trigonal-warping effect can be approximated by
\begin{equation}
\varphi_{\rm{eff}}^{\rm{trigonal}} = {\beta a_{c-c} \over 2d_{t}}q^{2} \cos(3\alpha),
\label{warping eqn}
\end{equation}
where $\varphi_{\rm{eff}}^{\rm{trigonal}}$ is the effective magnetic flux threading a nanotube equal in energy to the trigonal-warping splitting, $\beta$ is a constant of order unity, and all other quantities are defined as before.  As a note, the higher-order $k \cdot p$ formalism has been shown to be equivalent with tight-binding approach with regards to both the curvature-induced band gap and trigonal-warping effects.\cite{AjikiAndo96JPSJ, Ando04JPSJ, UryuAndo08PRB}  

\begin{figure}
\centering
\includegraphics[scale=.725]{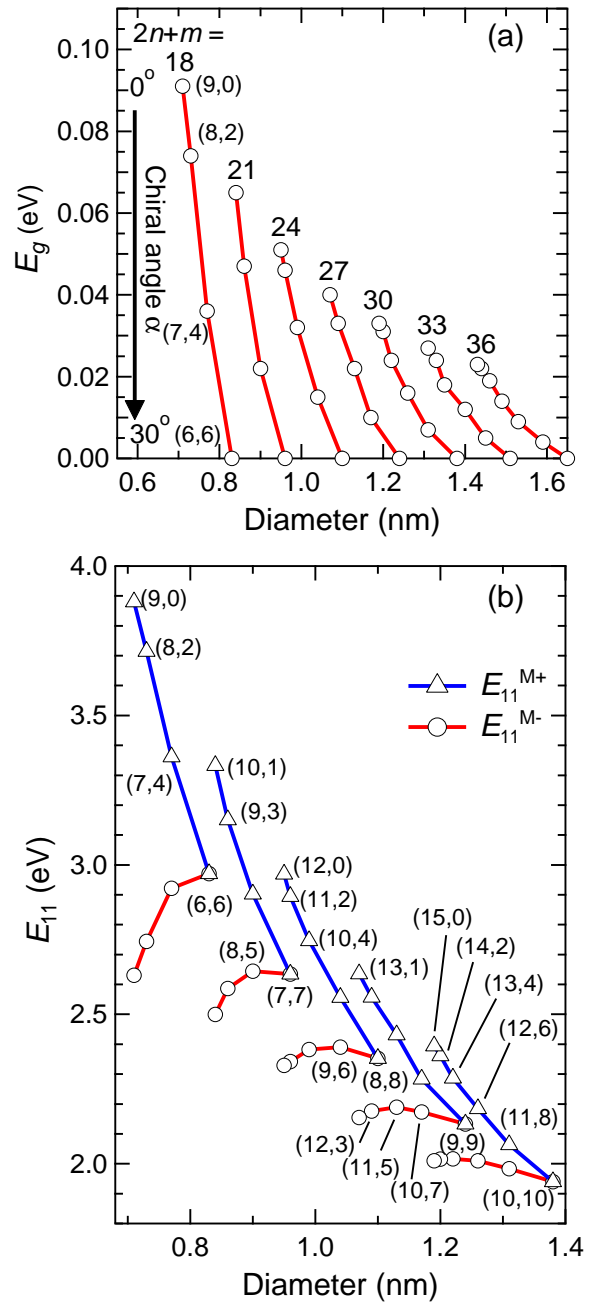}
\caption{Chirality dependence of the curvature-induced gap and trigonal-warping splitting in $\nu$~=~0 or ``metallic" SWCNTs.  (a)~A plot of calculated curvature-induced gap as a function of nanotube diameter with members of the same $2n$+$m$ family grouped together using Eq.~(\ref{curvature eqn}).  (b)~A plot of predicted $E_{11}^{M}$ optical transition energies as a function of nanotube diameter with members of the same $2n$+$m$ family grouped together for $\nu$~=~0 SWCNTs.\cite{NugrahaetAl10APL}  Upper-branch $E_{11}^{M+}$ transitions are labeled in blue and lower-branch $E_{11}^{M-}$ transitions are labeled in red.  The energy splitting between upper and lower branches is caused by trigonal warping.  Both curvature-induced gap and trigonal-warping splitting are zero for armchair species and maximum for ($3n,0$) zigzag species.}
\label{fig-warping}
\end{figure}

The trigonal-warping effect is most clearly shown in a comparison of the first massive bands ($|q|=1$) for armchair and non-armchair species in Fig.~\ref{fig-MetalBand}.  In Fig.~\ref{fig-MetalBand}a, a single parabolic band exists for each of the valence and conduction bands for the armchair case.  However, for the non-armchair case as shown in Fig.~\ref{fig-MetalBand}b, each parabolic band has split into two, a lower (1$^{-}$) and upper (1$^{+}$) branch, by trigonal warping.  The magnitude of the trigonal-warping splitting, $\Delta E_{ii}^{M}$, is chiral-angle-dependent, being 0 for armchair species and increasing with decreasing chiral angle up to its maximum value for ($3n,0$) zigzag species [see Eq.~(\ref{warping eqn})].  Additionally, $\Delta E_{ii}^{M}$ increases in magnitude with increasing electronic transition order, i.e., $\Delta E_{11}^{M} <  \Delta E_{22}^{M}$.  Using an extended tight-binding calculation, Fig.~\ref{fig-warping} and Tables~\ref{table-solution}-\ref{table-carpet2} provide theoretical estimates of the curvature-induced gap and trigonal-warping splitting for various $\nu$~=~0 species of SWCNTs.  Both effects were experimentally realized by scanning tunneling spectroscopy measurements on zigzag tubes, where a splitting of the van Hove singularities and gap opening at the Fermi energy were observed in the electronic density-of-states.\cite{OuyangetAl01Science}

\subsection{Selection Rules for Optical Transitions}
To properly predict, model, and interpret the optical properties of armchair and non-armchair $\nu$~=~0 SWCNTs in their respective spectra through such processes as optical absorption and Raman scattering, we must know which optical transitions are allowed between which energy states (i.e., optical selection rules).  Using the $k \cdot p$ method, Ajiki and Ando\cite{AjikiAndo94Physica} predicted the allowed optical transitions and optical conductivities for an armchair SWCNT.  Interband transitions between the massive, hyperbolic bands of the valence and conduction bands, with the same band index, $q$, are allowed if the polarization of the incident light is parallel to the nanotube axis.  However, interband optical transitions between the massless, linear bands are not allowed for any light polarization due to symmetry.  In the case where the polarization of the excitation is perpendicular to the nanotube axis, interband transitions are allowed only between bands where the band index changes by 1, i.e., $\Delta q = \pm1$ for both massive and massless bands.  However, due to the depolarization effect, these perpendicular transitions are expected to be heavily suppressed,\cite{AjikiAndo94Physica} although excitonic effects are predicted to retain this transition as a well-defined peak in absorption at a renormalized energy.\cite{UryuAndo06PRB}

Subsequently, Jiang \textit{et al.},\cite{JiangetAl04Carbon} using the tight-binding method, calculated the electric dipole matrix elements (whose square is proportional to optical absorption), yielding analytical expressions as a function of chiral index and wavevector, $k$.  As Ajiki and Ando, they found that parallel interband transitions are allowed only between massive bands of the same index and $k$-vector and not at all allowed between massless linear bands, for armchair species.  Furthermore, by examining the $k$-dependence of the matrix element, they found that the dipole matrix element reaches a maximum for $k$-values that coincide with the positions of the van Hove singularities (VHS) in the electronic density-of-states for each band.  As a result, the strong optical absorption observed in nanotubes is a result of the combination not only of the singularity in the density-of-states but also the coinciding maxima in dipole matrix elements.  Additionally for armchair SWCNTs, it was shown that the dipole matrix element is zero for all bands at the $k$~=~0 point, indicating a node in optical absorption for armchair species due to their high symmetry.  For polarization perpendicular to the tube axis, Jiang \textit{et al.}\cite{JiangetAl04Carbon} reached similar conclusions to that of Ajiki and Ando\cite{AjikiAndo94Physica} with the further refinement that only transitions between ``linear" and parabolic bands are significant in intensity as they are allowed everywhere in $k$-space.  Transitions between massive bands around $k$-points near the VHSs, however, are heavily suppressed due to the appearance of nodes in the dipole matrix elements at the positions of the VHS.

With the application of a magnetic field, a band gap opens, proportional in size to the magnetic flux, $\phi$.\cite{AjikiAndo93JPSJ, AjikiAndo94Physica, KonoRoche06CRC, KonoetAl07Book} The band gap reaches a maximum when $\phi = \phi_0 / 2$ (where $\phi_0 = h/e$ is the magnetic flux quantum) and then decreases to zero at $\phi=\phi_{0}$.  This gap opening allows optical transitions between the previously massless linear bands (which are now massive in the vicinity of the band gap) for light polarized parallel to the tube axis.  Effectively, the magnetic field has changed a metallic armchair nanotube into a narrow-gap semiconductor.  All other selection rules previously mentioned remain the same.  As a result, the optical selection rules for narrow-gap semiconductors (non-armchair, $\nu$~=~0 species) in the absence of a magnetic field are the same as for armchair species in the presence of a magnetic field.  

With regards to resonant Raman processes, the optical selection rules are similar to those of optical absorption with the added considerations of the inclusion of the emission or absorption of a phonon of a given symmetry and the emission of the scattered photon.  For first-order Raman processes, corresponding to vibrational modes such as the radial breathing mode (RBM)  ($A$-symmetry) and the G-band ($A$-, $E_{1}$- and $E_{2}$-symmetry components), $A$- and $E_{1}$-symmetry phonons can be observed between states $E_{q} \rightarrow E_{q}$ (parallel interband transition $E_{ii}$) and $E_{q} \rightarrow E_{q \pm 1}$ (perpendicular interband transition $E_{ij}$ where $j=i \pm1$), where $q$ is the index identifying a particular VHS in the valence [before (after) arrow] and conduction [after (before) arrow] bands.\cite{JorioetAl03PRL}  $E_{2}$-symmetry phonons can only be observed via the $E_{q} \rightarrow E_{q \pm 1}$ excitation process.  Because of the strong suppression of perpendicular transitions due to the depolarization effect,\cite{AjikiAndo94Physica} phonons excited via $E_{ii}$ typically dominate Raman spectra of ensemble samples.

\subsection{Excitonic Effects}
An important additional factor to consider in the optical spectra of metallic carbon nanotubes is the role excitons play.  While excitons are commonly associated with the optical transitions and spectra of semiconductors, recent experiments\cite{WangetAl07PRL2, Doornetal08PRB, TuetAl11JACS, HarozetAl12JACS} and theoretical studies\cite{Ando97JPSJ, DeslippeetAl07NL, UryuAndo08PRB, MalicetAl10PRB, HartmannetAl11PRB}  have reported the strong influence of excitonic effects in the optical spectra and band structure of metallic SWCNTs.  While the optical selection rules governing absorption and resonant Raman processes remain unaffected,\cite{UryuAndo08PRB} the energy positions and line shapes of the major optical features are modified as a result of the formation of excitons.  Allowed optical transitions for excitons associated with a particular electronic state still obey the earlier optical selection rules with the additional constraint that only 1 of the 16 possible excitonic states (after taking into account valley and spin degeneracies) is optically active.\cite{Ando06JPSJ}  Therefore, optical transitions are only allowed for such optically active excitonic states in the absence of additional perturbations.  

The inclusion of excitonic effects into the optical spectra of $\nu$~=~0 SWCNTs decreases the optical transition energy by the exciton binding energy, estimated to be $\sim$50-100~meV for the first optical transition in ``metallic" species\cite{DeslippeetAl07NL, UryuAndo08PRB, MalicetAl10PRB} with diameters ranging from 0.8 to 2~nm, with binding energy decreasing with increasing nanotube diameter.  Furthermore, the electron-hole Coulomb interaction causes a significant suppression of the electronic continuum in 1-D systems.\cite{OgawaTakagahara91PRB, OgawaTakagahara91PRBRC}  The result of the continuum suppression on optical spectra is a significant decrease in the high-energy tail in absorption due to the VHS.  This causes the line-shape of absorption features to become highly symmetric and almost Lorentzian-like, but a small high-energy tail still remains.\cite{DeslippeetAl07NL, UryuAndo08PRB, MalicetAl10PRB}  This is due to the incomplete transfer of oscillator strength from the continuum to the excitonic states.\cite{MalicetAl10PRB}  Interestingly, predictions by Malic \textit{et al.}\cite{MalicetAl10PRB} show that the magnitude of the trigonal-warping-induced splitting of the $E_{11}^{M}$ optical transition into $E_{11}^{M-}$ and $E_{11}^{M+}$ is not significantly affected by excitonic effects.  This should produce observable absorption features in non-armchair, $\nu$~=~0 species of sufficiently small diameter where the trigonal-warping splitting can exceed the line-widths.  Low-energy excitation of excitons across the curvature-induced band gap in non-armchair $\nu$~=~0 SWCNTs or the magnetic-field-opened band gap in armchair SWCNTs should produce observable optical absorption in the terahertz frequency range only in the local $k$-space vicinity of the gap.  Both lowest lying excitons (one optically inactive and one optically active) will be equally populated at experimentally attainable temperatures removing any suppression of optical emission.\cite{HartmannetAl11PRB}  Excitations away from the band gap, however, will remain forbidden.\cite{UryuAndo08PRB, HartmannetAl11PRB}  No excitonic phases are predicted to exist at any value of band gap, since the exciton binding energy monotonically decreases with decreasing band gap.\cite{Ando97JPSJ,HartmannetAl11PRB}

\section{Previous Spectroscopic Results on $\nu=0$ Carbon Nanotubes}
\label{previous}

Optical studies of metallic carbon nanotubes have evolved in richness with the significant progress in sample preparation and availability.  Initial studies on powders and films of as-produced SWCNT material, first produced in the 1990s, confirmed general predictions of the optical and electronic properties of SWCNTs.  Later, in the early 2000s, with the development of aqueous surfactant suspensions containing highly individualized tubes using ultrasonication and ultracentrifugation,\cite{OConnelletAl02Science} our ability to study optical and electronic properties of specific ($n,m$) species grew, allowing researchers to refine early models and theories.  Most recently, with the advent of highly selective separation techniques such as density gradient ultracentrifugation (DGU) and DNA-based ion-exchange chromatography, optical studies of single-chirality SWCNT samples on macroscopic scales were possible, which are devoid of the influence of other neighboring and often interfering species.  At the same time, advancements in single-molecule, micro-spectroscopies and low-density, long-length SWCNT growth have allowed researchers to also probe properties of single ($n,m$) nanotube species.  Presented here is a brief summary of the important contributions to the current literature surrounding the properties of $\nu$~=~0 SWCNTs across the techniques of optical absorption, resonant Raman scattering, and resonant Rayleigh scattering spectroscopies.  

\subsection{Optical Absorption}

Kataura \textit{et al.}\cite{KatauraetAl99SM} performed one of the earlier optical absorption studies on thin films that were produced from ethanol dispersions of SWCNTs synthesized by the electric arc-discharge method.  The ethanol dispersion was spray-painted onto a quartz substrate and optical absorption was measured in the $\approx$2500-200~nm range.  Although the sample was composed of aggregates (bundles) of SWCNTs, three distinct, broad absorption features were observed corresponding to the first and second optical transitions of semiconducting SWCNTs and the first optical transition of ``metallic" or $\nu$~=~0 SWCNTs, confirming predictions originally made by nearest-neighbor tight-binding approximation calculations.  However, absorption features could not be attributed to specific chiral structures.  More importantly, this study marks the first appearance of what would later be named the so-called Kataura plot, a plot of optical interband transitions versus nanotube diameter.  Such a plot has widely been used by subsequent researchers to aid in the assignment of measured optical transitions to specific ($n,m$) species and has served a predictive tool as well.  

After this initial work, further information obtained from optical absorption spectroscopy progressed slowly due to limitations imposed by sample quality.  Although liquid suspensions of nanotubes could be produced via ultrasonication and/or chemical functionalization, absorption spectra remained poorly defined due to the reaggregation of SWCNTs in suspension or modification of band structure due to functionalization, producing results similar to the initial work by Kataura \textit{et al.}  In 2002, O'Connell \textit{et al.},\cite{OConnelletAl02Science} however, succeeded in suspending SWCNTs in water using surfactant micelles and a combination of ultrasonication and ultracentrifugation using nanotubes produced by the high-pressure carbon monoxide (HiPco) method.  This breakthrough resulted in aqueous suspensions containing individualized SWCNTs that were not significantly modified electronically by the suspension process.  Optical absorption spectra of such samples revealed a multitude of optical features corresponding to the distinct optical transitions of semiconducting and metallic species.  Furthermore, direct band gap fluorescence was observed for semiconducting species in the near-infrared region in this study.  Although such a photoluminescence process does not occur for metallic nanotubes, scanning excitation studies did reveal a region in the visible excitation ($\sim$450-550~nm) where virtually no photoluminescence was observed, corresponding to the optical transitions of metallic nanotubes.  Most importantly, these types of aqueous surfactant suspensions of SWCNTs opened the door to many other optical studies via absorption, resonant Raman scattering, and ultrafast optical studies.  As related to optical absorption studies, these solution-phase samples along with subsequent ($n,m$) identification of optical features by other techniques, allowed studies of chemical doping and functionalization (for example, diazonium functionalization chemistry by Strano \textit{et al.}\cite{StranoetAl03Science}).  Such studies revealed that, in many cases, metallic nanotubes were more susceptible to such chemical species presumably because of the facile availability of delocalized $\pi$-electrons.  

With the rapid progress in separation techniques such as DGU that resulted in samples that were separated by electronic type and/or diameter, clear delineation of semiconducting and metallic transitions could be observed.\cite{ArnoldetAl06NatureNano, YanagietAl08APE}  In particular,  with the knowledge of the general excitation regions where metallic and semiconducting transitions occur, optical absorption was used to estimate the electronic type purity of enriched samples by integrating the area under metallic and semiconducting species.  Even more useful was the observation by Miyata \textit{et al.}\cite{MiyataetAl08JPCC} and Green \textit{et al.}\cite{GreenHersam08NL} that thin films produced from metal-enriched nanotube material exhibit little broadening and red-shifting as compared to the individualized enriched suspension.  This is unlike the case of semiconductor-enriched suspensions, which exhibit significant peak broadening and red-shifting upon aggregation of SWCNTs into films.  Using length-sorted samples produced via DGU from CoMoCAT SWCNT material, Searles \textit{et al.}\cite{SearlesetAl10PRL} performed magneto-optical-absorption studies in magnetic fields up to 35~T, where they determined the orbital magnetic susceptibility anisotropy, $\Delta \chi$, to be 2-4 times greater for $\nu$~=~0 species (5,5), (6,6) and (7,4) than for semiconducting species of comparable diameter.

At the same time that sample preparation improved, absorption spectroscopy techniques improved in sensitivity to allow the measurement of absorption spectra of individual SWCNTs.  Berciaud \textit{et al.}\cite{BerciaudetAl07NL} succeeded in measuring absorption spectra, using photothermal heterodyne imaging (PHI) and a cw tunable laser excitation source, of individual HiPco-derived SWCNTs, deposited from aqueous suspension onto a microscope coverslip.  Spots observed in PHI images were considered to be individual tubes and had subsequent absorption and photoluminescence spectra measured.  In cases where both absorption and photoluminescence were observed, the tubes were attributed to semiconducting species.  Such nanotubes also exhibited exciton-phonon sidebands in their spectra, additionally confirming their semiconducting nature.  In cases where PHI spots only produced an absorption spectrum and no excition-phonon sidebands, such tubes were attributed to metallic species although assignment to a specific ($n,m$) species was not possible.  Wang \textit{et al.},\cite{WangetAl07PRL2} using SWCNTs grown by the chemical vapor deposition (CVD) method across a trench produced in a support substrate, measured optical absorption spectra using a modulation technique and broadband white light source produced by supercontinuum generation.  Individual SWCNTs were independently ($n,m$) identified by electron beam diffraction before absorption measurement.  In particular, Wang \textit{et al.}~measured the absorption spectrum of a (21,21) armchair nanotube and observed a highly symmetric absorption line shape followed by a very weak continuum at slightly higher energy.  This highly symmetric feature was attributed to excitonic absorption and constituted the first experimental evidence that excitons can exist in one-dimensional carbon nanotube metals (first predicted by Deslippe \textit{et al}).\cite{DeslippeetAl07NL}  Additionally, Wang \textit{et al.} and later Zeng \textit{et al.},\cite{ZengetAl09PRL} using Rayleigh scattering in a backscattering geometry on an individual (13,10), claimed to observe a weak exciton-phonon sideband due to an optical phonon, which supports the claim of exciton involvement in metallic SWCNT optics.  This feature, however, was not clearly observed in either case.

\subsection{Resonant Raman Scattering}
Resonant Raman scattering (RRS) was one of the earliest optical techniques to gather information specifically surrounding metallic carbon nanotubes.  Because of Raman scattering's resonant nature and the unique band structure of carbon nanotubes, only phonons corresponding to species electronically resonant with the excitation energy should appear.  Combined with the frequency dependence of certain vibrational modes for nanotube crystal structure, ($n,m$)-specific information can be extracted.

The first RRS spectra measured on carbon nanotubes were reported by Rao \textit{et al.}\cite{RaoetAl97Science} on laser-ablation-synthesized SWCNT powder.  By examining excitation energy dependence using multiple discrete laser wavelengths, multiple Raman-active modes predicted by theory were observed including the radial breathing mode (RBM) and G-band mode.  The RBM and G-band modes were in particular observed to change with excitation energy, illustrating the resonance effect in nanotube Raman scattering.  At one particular excitation wavelength, the G-band was observed to become anomalously broad, which would later be shown to be due to a unique feature of non-armchair, $\nu$~=~0 nanotubes.\cite{WuetAl07PRL2, MicheletAl09PRB, HarozetAl11PRB}  To examine this resonant effect more closely, Sugano \textit{et al.}\cite{SuganoetAl98CPL} performed more detailed excitation-wavelength-dependent RRS studies using a tunable excitation source (cw dye laser) on SWCNT powder produced by the electric arc-discharge method.  There they observed resonant enhancement of certain RBM features that seemed to agree with theoretical predictions of the locations of electronic transitions for certain tubes.  Furthermore, the predicted diameter dependence of electronic transitions was observed, with transition energy decreasing with decreasing RBM frequency (i.e., increasing nanotube diameter).  

Moving focus to the G-band phonon, Pimenta  \textit{et al.}\cite{PimentaetAl98PRB} made the association of the broad G-band feature observed by Rao \textit{et al.} to ``metallic" SWCNTs using laser-ablation-produced SWCNTs.  Brown \textit{et al.}\cite{BrownetAl01PRB} subsequently assigned the broad, lower frequency component of the G-band, the so-called G$^{-}$ peak, to a Breit-Wigner-Fano resonance between the discrete circumferential vibrational mode and the continuum of electronic states.  As such, the broad G$^{-}$ peak was used as an indicator of the presence of metallic nanotubes for many years until further refinements were made (see Section \ref{RRS-Gband}).  In another work by Brown \textit{et al.},\cite{BrownetAl00PRB} it was shown that anti-Stokes and Stokes Raman spectra taken at the same excitation wavelengths demonstrated very different line shapes in their respective G-bands, one corresponding to metallic species and the other to semiconducting species.  This demonstrated the discrete nature of the electronic density-of-states of a given species and the ability to selectively excite one species over another by careful selection of laser energy.  

However, in mixed powder samples like those employed above, very little ($n,m$)-specific data can be obtained.  To resolve this issue, researchers resorted to the use of RRS micro-spectroscopy to investigate individual SWCNTs on a supporting substrate.  Using such an approach, Jorio \textit{et al.}\cite{JorioetAl01PRL} measured Raman spectra of multiple individual tubes, grown via CVD onto silicon substrates, focusing on the RBM frequency.  Combined with predictions based on a theoretical Kataura plot, RBMs were assigned to possible ($n,m$) indices, including several metallic species, providing definitive assignments of Raman spectral features to specific metallic chiralities.  Jorio \textit{et al.}\cite{JorioetAl02PRB} followed up their earlier work with a study of the G-band frequencies of 62 different individual tubes.  With that array of data on different ($n,m$)-identified species, they attempted to extract an empirical relationship describing the frequency dependence of the lower energy G$^{-}$ peak relative to the G$^{+}$ peak, which was claimed to have no frequency dependence ($\omega_{G_{-}}=1590$~cm$^{-1}$$-C/d_{t}^{2}$, where $C$ is a constant dependent on electronic type).  

Examining yet another phonon mode, the dispersive G'-mode, Souza Filho \textit{et al.}\cite{SouzaFilhoetAl02CPL} measured the intensity of that mode as a function of excitation energy for several individual metallic SWCNTs grown via CVD on SiO$_{2}$.  The observed dispersion was claimed to correlate with the then-recently-predicted trigonal warping effect in the electronic density-of-states.  Furthering investigations on the trigonal warping effect via Raman, Son \textit{et al.}\cite{SonetAl06PRB} observed both lower- ($E_{ii}^{M-}$) and upper-branch ($E_{ii}^{M+}$) metallic optical transitions, due to trigonal warping splitting (see Figs. \ref{fig-MetalBand}b and \ref{fig-warping}b), in individual metallic tubes via measurement of the intensity dependence of a tube's RBM as a function of excitation energy (Raman excitation profile).   There, it was observed that the higher-energy, upper-branch metallic transition resulted in lower Raman intensity at its resonance maximum as compared to the lower energy, lower-branch transition for the same tube.  However, this only occurred for nanotubes of large diameter ($>$1.5~nm) causing questions as to why its appearance had not previously been observed in smaller diameter tubes.  

Returning to the G-band mode, experimental studies by Wu \textit{et al.},\cite{WuetAl07PRL} Telg \textit{et al.},\cite{TelgetAl08PSSB} Fouquet \textit{et al.},\cite{FouquetetAl09PRL} Michel \textit{et al.},\cite{MicheletAl09PRB} and Park \textit{et al.}\cite{ParketAl09PRB}~attempted to correlate G-band line shape with specific ($n,m$), especially in light of a new explanation for the origin of the broad G$^{-}$ feature in metallic nanotubes as arising from the Kohn anomaly\cite{PiscanecetAl07PRB} (discussed later in Section \ref{RRS-Gband}).  Raman measurements were made on individual, ($n,m$)-identified tubes.  In all the studies except Park \textit{et al.}, the appearance of the broad G$^{-}$~peak was attributed to a $\nu$~=~0, non-armchair species with the largest G$^{-}$ intensity observed for zigzag-like species.  In the case of armchair nanotubes, the G$^{-}$~peak was absent.  Park \textit{et al.}'s observations agreed with that of the other reports mentioned prior except in the case of the armchair SWCNT, where they claimed a broad G$^{-}$~peak was present as well.  In a completely new avenue of investigation, Farhat \textit{et al.}\cite{FarhatetAl11PRL} observed a broad Raman feature not corresponding to previously observed vibrational Raman features for individual $\nu$~=~0 tubes on a substrate.  The broad Raman feature, which moved in scattered photon energy with changing excitation energy, was explained to be a result of inelastic scattering of photoexcited carriers with the electronic continuum present in the linear bands of metallic band structure.  However, it should be pointed out that the species examined in this study were not truly metallic as they were non-armchair, $\nu$~=~0 nanotubes. 

One particular issue with single-tube Raman measurements is the inability to select which ($n,m$) are investigated due to the lack of control in the growth process.  Finding a particular type of chirality of nanotube is a matter of chance.  With the development of aqueous surfactant suspensions, however, all species present in the as-produced nanotube material can be examined if the appropriate excitation conditions are available.  Strano \textit{et al.},\cite{StranoetAl03NL} using a surfactant-suspended dispersion of HiPco SWCNTs, attempted to use a tunable excitation source (a combination of cw dye and solid-state lasers) to map out most of the metallic species via excitation dependence of their RBM features and assign them to specific ($n,m$) indices.  While not all the assignments were correct due to the unknown knowledge that upper branch metallic transitions were not present in nanotubes with a diameter $<$1.3~nm, the basic methodology to assigning them was sound.  Subsequent studies by Doorn \textit{et al.},\cite{DoornetAl04APA} Telg \textit{et al.},\cite{TelgetAl04PRL} Fantini \textit{et al.},\cite{FantinietAl04PRL} and Maultzsch \textit{et al.}\cite{MaultzschetAl05PRB} on surfactant-suspended HiPco nanotubes and later surfactant-suspended CoMoCAT nanotubes,\cite{FantinietAl07CPL} correctly assigned RBM features using optical transition energies obtained from Raman excitation profiles to ($n,m$) species.  The combined data sets of tabulated ($n,m$) with RBM frequencies and optical transition energies allowed Maultzsch \textit{et al.}\cite{MaultzschetAl05PRB} and also Jorio \textit{et al.}\cite{JorioetAl05PRB} to develop empirical equations to predict both RBM frequencies and optical transition energies for nanotubes, which has served as an experimental predictive tool for locating previously unobserved spectral features.\cite{NanotetAl12AM}  Jorio \textit{et al.},\cite{JorioetAl05PRB} in particular, attempted to take existing expressions for tight-binding calculations predicting SWCNT optical transitions and add additional chiral-angle-dependent and logarithmic terms to account for chiral-dependent electron and hole effective masses and many-body corrections, respectively, with parameters tuned to fit existing optical data.  

Looking at another type of ensemble nanotube sample, arrays of vertically aligned carbon nanotubes grown via CVD and containing a broad range of diameters and chiralities, Doorn \textit{et al.}\cite{Doornetal08PRB} performed similar excitation-wavelength-dependent studies to that of the tunable excitation suspension work previously mentioned.  Due to the broad diameter distribution (0.7-4~nm), 77 different metallic optical transitions were observed including 15 clear, upper-branch metallic transitions (the rest were lower-branch metallic transitions), spanning the first and second optical transitions.  The data show that upper-branch metallic optical transitions were not observed for SWCNTs with a diameter $<$1.3~nm, yet for diameters $>$2~nm, significant Raman intensity due to resonance with upper-branch transitions was observed, especially for zigzag and near-zigzag species, in contrast to the work of Son \textit{et al.}\cite{SonetAl06PRB}  By examining scaling behavior of the optical transitions to a model previously created for examining semiconducting nanotube transitions\cite{AraujoetAlPRL2007} where excitons are known to play an important role, Doorn \textit{et al.}~suggested that excitons are also important in the optical transitions of metallic nanotubes, in agreement with single-tube absorption measurements.\cite{WangetAl07PRL}  To further investigate this connection between excitons and $\nu$~=~0 nanotubes, May \textit{et al.},\cite{MayetAl10PRB} using similar vertically aligned carbon nanotube arrays, examined the optical transition energy dependence of the metallic species (13,1) via its RBM as a function of temperature.  At moderate temperatures above room temperature, the optical transition energy was observed to decrease with increasing temperature.  This monotonic decrease in transition energy up to 570~K was explained as a softening of the optical transition due to lattice expansion and electron-phonon coupling upon heating.  However, above 570~K, a sudden discontinuity and increase occurred in the optical transition energy with temperature.  This was attributed to thermal dissociation of the exciton bound state (exciton binding energies for metallic nanotubes have been estimated to be $\approx$~50~meV,\cite{DeslippeetAl07NL, WangetAl07PRL} which corresponds to 580~K) with the optical transition state now consisting of band-to-band transitions.  This work and the previous one\cite{Doornetal08PRB} highlight the importance of excitons in metallic nanotubes, as will be discussed later in Section \ref{absorption}.

Tables \ref{table-solution}-\ref{table-carpet2} summarize many of the basic ($n,m$)-specific nanotube parameters and experimentally determined data from optical absorption and resonant Raman experiments reviewed briefly here and presented later on in Sections \ref{absorption}-\ref{RRS-RBM}, along with theoretical predictions of curvature-induced band gap and trigonal warping splitting.

\begin{table*}
\small
  \caption{\ Basic parameters, optical transition energies, and phonon frequencies of $\nu = 0$ (or mod 0), single-wall carbon nanotubes, suspended in aqueous surfactant with a diameter range from 0.73 to 1.65~nm.  Here, ($n,m$) are the chiral indices, $2n$+$m$ is the family number,  $d_{t}$ is the nanotube diameter, RBM is the radial breathing mode frequency, $\alpha$ is the chiral angle, $E_{g}$ is the curvature-induced band gap, $\Delta E_{11}^{M}$ is the trigonal warping splitting =~$E_{11}^{M+}$$-$$E_{11}^{M-}$, and $E_{11}^{M-}$ is the lower-branch optical transition.  $^{\dagger}$Values calculated using Eq.~(\ref{curvature eqn}).\cite{KaneMele97PRL}  $^{*}$Values determined from extended tight-binding calculations of $E_{11}^{M}$ (see Fig.~\ref{fig-warping}b) taken from Reference \onlinecite{JiangetAl07PRB}.  $^{a}$Values determined from RRS measurements in this work for armchair-enriched suspension of HiPco SWCNTs.  $^{b}$Values determined from RRS and optical absorption measurements on DNA-based-method-separated single-chirality suspension of HiPco SWCNTs.\cite{TuetAl11JACS}  $^{c}$Values determined from RRS measurements in this work for sodium deoxycholate (DOC) suspension of HiPco SWCNTs.  $^{d}$RBM values taken from RRS measurements in this work for armchair-enriched suspension of laser-ablation SWCNTs; $E_{11}^{M-}$ estimated from optical absorption.  $^{e}$RBM value taken from RRS measurements in this work for armchair-enriched suspension of arc-discharge SWCNTs; $E_{11}^{M-}$ estimated from optical absorption. $^{\ddagger}$Values taken from RRS measurements in Reference \onlinecite{MaultzschetAl05PRB} for sodium dodecyl sulfate (SDS) suspension of HiPco SWCNTs.  $^{\triangleright}$Values taken from RRS measurements in Reference \onlinecite{FantinietAl04PRL} for sodium dodecyl sulfate (SDS) suspension of HiPco SWCNTs. $^{\star}$RBM observed at a single excitation energy; $E_{11}^{M-}$ could not be determined.}
  \label{table-solution}
  \begin{tabular*}{\textwidth}{@{\extracolsep{\fill}}llllllll}
    \hline
   ($n,m$)	&	$2n$+$m$	&	$d_{t}$ (nm)	&	RBM (cm$^{-1}$)	&	$\alpha$ (deg)	&	$E_{g}$ (eV)$^{\dagger}$	&	$\Delta E_{11}^{M}$ (eV)$^{*}$ 	&	$E_{11L}^{M}$ (eV) \\
\hline
(6,6)	&	18	&	0.83	&	285,$^{a}$	289,$^{b}$	286$^{c}$						&	30.0	&	0	&	0.00	&	2.74,$^{a}$	2.71,$^{b}$	2.71$^{c}$								\\
(7,4)	&	18	&	0.77	&							304,$^{c}$	305$^{\ddagger}$			&	21.1	&	36	&	0.44	&							2.65,$^{c}$	2.63$^{\ddagger}$					\\
(8,2)	&	18	&	0.73	&							312,$^{c}$	316,$^{\ddagger}$			&	10.9	&	74	&	0.97	&							2.49,$^{c}$	2.52$^{\ddagger}$					\\
(7,7)	&	21	&	0.96	&	244,$^{a}$	250,$^{b}$	246,$^{c}$	248,$^{\ddagger}$	248$^{\triangleright}$	&	30.0	&	0	&	0.00	&	2.48,$^{a}$	2.46,$^{b}$	2.48,$^{c}$	2.45,$^{\ddagger}$	2.43$^{\triangleright}$			\\
(8,5)	&	21	&	0.90	&	261,$^{a}$				260,$^{c}$	263,$^{\ddagger}$	264$^{\triangleright}$	&	22.4	&	22	&	0.26	&	2.49,$^{a}$				2.48,$^{c}$	2.47,$^{\ddagger}$	2.43$^{\triangleright}$			\\
(9,3)	&	21	&	0.86	&							269,$^{c}$	273,$^{\ddagger}$	274$^{\triangleright}$	&	13.9	&	47	&	0.56	&							$^{\star}$,$^{c}$	2.43,$^{\ddagger}$	2.35$^{\triangleright}$			\\
(10,1)	&	21	&	0.84	&										276$^{\ddagger}$			&	4.7	&	65	&	0.83	&										2.38$^{\ddagger}$					\\
(8,8)	&	24	&	1.10	&	217,$^{a}$				216,$^{c}$				218.5$^{\triangleright}$	&	30.0	&	0	&	0.00	&	2.24,$^{a}$				2.26,$^{c}$				2.22$^{\triangleright}$			\\
(9,6)	&	24	&	1.04	&	229,$^{a}$				229,$^{c}$				230$^{\triangleright}$	&	23.4	&	15	&	0.17	&	2.25,$^{a}$				2.23,$^{c}$				2.24$^{\triangleright}$			\\
(10,4)	&	24	&	0.99	&							238,$^{c}$				239.2$^{\triangleright}$	&	16.1	&	32	&	0.36	&							2.22,$^{c}$				2.22$^{\triangleright}$			\\
(11,2)	&	24	&	0.96	&							241,$^{c}$				244.4$^{\triangleright}$	&	8.2	&	46	&	0.55	&							2.17,$^{c}$				2.19$^{\triangleright}$			\\
(12,0)	&	24	&	0.95	&										245,$^{\ddagger}$	247$^{\triangleright}$	&	0.0	&	51	&	0.64	&										2.18,$^{\ddagger}$	2.16$^{\triangleright}$			\\
(9,9)	&	27	&	1.24	&	192,$^{a}$				192,$^{c}$	195,$^{\ddagger}$	196.4$^{\triangleright}$	&	30.0	&	0	&	0.00	&	2.06,$^{a}$				2.06,$^{c}$	2.02,$^{\ddagger}$	2.03$^{\triangleright}$			\\
(10,7)	&	27	&	1.17	&	201,$^{a}$				201,$^{c}$	204,$^{\ddagger}$	205.6$^{\triangleright}$	&	24.2	&	10	&	0.11	&	2.10,$^{a}$				2.09,$^{c}$	2.07,$^{\ddagger}$	2.07$^{\triangleright}$			\\
(11,5)	&	27	&	1.13	&	210,$^{a}$				209,$^{c}$	212,$^{\ddagger}$	214$^{\triangleright}$	&	17.8	&	22	&	0.24	&	2.08,$^{a}$				2.09,$^{c}$	2.08,$^{\ddagger}$	2.06$^{\triangleright}$			\\
(12,3)	&	27	&	1.09	&							214,$^{c}$	217,$^{\ddagger}$	219$^{\triangleright}$	&	10.9	&	33	&	0.38	&							2.06,$^{c}$	2.08,$^{\ddagger}$	2.04$^{\triangleright}$			\\
(13,1)	&	27	&	1.07	&										220,$^{\ddagger}$	223$^{\triangleright}$	&	3.7	&	40	&	0.48	&										2.06,$^{\ddagger}$	2.02$^{\triangleright}$			\\
(10,10)	&	30	&	1.38	&	174,$^{a}$				174,$^{c}$	176$^{\ddagger}$			&	30.0	&	0	&	0.00	&	1.90,$^{a}$				1.89,$^{c}$	1.89$^{\ddagger}$					\\
(11,8)	&	30	&	1.31	&	182,$^{a}$				182,$^{c}$	183,$^{\ddagger}$	185.4$^{\triangleright}$	&	24.8	&	7	&	0.08	&	1.93,$^{a}$				1.92,$^{c}$	1.94,$^{\ddagger}$	1.90$^{\triangleright}$			\\
(12,6)	&	30	&	1.26	&	188,$^{a}$				189,$^{c}$	189,$^{\ddagger}$	191.6$^{\triangleright}$	&	19.1	&	16	&	0.17	&	1.94,$^{a}$				1.94,$^{c}$	1.95,$^{\ddagger}$	1.92$^{\triangleright}$			\\
(13,4)	&	30	&	1.22	&							194,$^{c}$	194,$^{\ddagger}$	196.5$^{\triangleright}$	&	13.0	&	24	&	0.27	&							1.92,$^{c}$	1.94,$^{\ddagger}$	1.93$^{\triangleright}$			\\
(14,2)	&	30	&	1.20	&										196,$^{\ddagger}$	200.5$^{\triangleright}$	&	6.6	&	31	&	0.35	&										1.93,$^{\ddagger}$	1.92$^{\triangleright}$			\\
(15,0)	&	30	&	1.19	&										200,$^{\ddagger}$	204.6$^{\triangleright}$	&	0.0	&	33	&	0.38	&										1.91,$^{\ddagger}$	1.88$^{\triangleright}$			\\
(11,11)	&	33	&	1.51	&	164.9$^{d}$												&	30.0	&	0	&	0.00	&															1.79$^{d}$	\\
(12,9)	&	33	&	1.45	&	169$^{d}$												&	25.3	&	5	&	0.06	&															1.79$^{d}$	\\
(13,7)	&	33	&	1.40	&	175.3$^{d}$												&	20.2	&	12	&	0.13	&															1.79$^{d}$	\\
(14,5)	&	33	&	1.35	&										175$^{\ddagger}$			&	14.7	&	18	&	0.20	&										1.83$^{\ddagger}$					\\
(15,3)	&	33	&	1.33	&										179$^{\ddagger}$			&	8.9	&	24	&	0.27	&										1.83$^{\ddagger}$					\\
(16,1)	&	33	&	1.31	&										182$^{\ddagger}$			&	3.0	&	27	&	0.31	&										1.81$^{\ddagger}$					\\
(12,12)	&	36	&	1.65	&	151.3$^{e}$												&	30.0	&	0	&	0.00	&															1.65$^{e}$	\\
 \hline
  \end{tabular*}
\end{table*}

\begin{table*}
\small
  \caption{\ Basic parameters, optical transition energies, and phonon frequencies of $\nu = 0$ (or mod 0), single-wall carbon nanotubes, produced as a vertically aligned array with a diameter range from 1.07 to 2.21~nm.   Here, ($n,m$), $2n$+$m$,  $d_{t}$, RBM, $\alpha$, and $E_{g}$ are defined as before.  $\Delta E_{ii}^{M}$ is the trigonal warping splitting of the $i$th optical transition =~$E_{ii}^{M+}$$-$$E_{ii}^{M-}$ and can be assumed to be $\Delta E_{11}^{M}$ unless otherwise indicated, $E_{11}^{M-}$ is the lower-branch of the first optical transition, $E_{11}^{M+}$ is the upper-branch of the first optical transition, $E_{22}^{M-}$ is the lower-branch of the second optical transition, and $E_{22}^{M+}$ is the upper-branch of the second optical transition.  $^{\dagger}$Values calculated using Eq.~(\ref{curvature eqn}).\cite{KaneMele97PRL}  $^{*}$Values determined from extended tight-binding calculations of $E_{11}^{M}$ and $E_{22}^{M}$ taken from Reference \onlinecite{JiangetAl07PRB}. RBM frequencies and $E_{ii}$ values taken from Reference \onlinecite{Doornetal08PRB}.  $^{a}$Calculated trigonal warping splitting is for  $\Delta E_{22}^{M}$.}
  \label{table-carpet1}
  \begin{tabular*}{\textwidth}{@{\extracolsep{\fill}}lllllllllll}
    \hline
   ($n,m$)	&	$2n$+$m$	&	$d_{t}$ (nm)	&	RBM (cm$^{-1}$)	&	$\alpha$ (deg)	&	$E_{g}$ (eV)$^{\dagger}$	&	$\Delta E_{ii}^{M}$ (eV)$^{*}$	&	$E_{11}^{M-}$ (eV)	&	$E_{11}^{M+}$ (eV)	&	$E_{22}^{M-}$ (eV)	&	$E_{22}^{M+}$ (eV)	\\
    \hline
(10,7)	&	27	&	1.17	&	197	&	24.2	&	10	&	0.11				&	2.1	&	-	&		&		\\
(11,5)	&	27	&	1.13	&	206	&	17.8	&	22	&	0.24				&	2.11	&		&		&		\\
(12,3)	&	27	&	1.09	&	211	&	10.9	&	33	&	0.38				&	2.112	&		&		&		\\
(13,1)	&	27	&	1.07	&	215	&	3.7	&	40	&	0.48				&	2.095	&		&		&		\\
(10,10)	&	30	&	1.38	&	166	&	30.0	&	0	&	0.00				&	1.88	&		&		&		\\
(11,8)	&	30	&	1.31	&	177	&	24.8	&	7	&	0.08				&	1.94	&		&		&		\\
(12,6)	&	30	&	1.26	&	184	&	19.1	&	16	&	0.17				&	1.967	&		&		&		\\
(13,4)	&	30	&	1.22	&	189	&	13.0	&	24	&	0.27				&	1.969	&		&		&		\\
(15,0)	&	30	&	1.19	&	193	&	0.0	&	33	&	0.38				&	1.977	&		&		&		\\
(14,5)	&	33	&	1.35	&	169	&	14.7	&	18	&	0.20				&	1.857	&		&		&		\\
(16,1)	&	33	&	1.31	&	175	&	3.0	&	27	&	0.31				&	1.867	&	2.111	&		&		\\
(14,8)	&	36	&	1.53	&	151	&	21.1	&	9	&	0.10				&	1.704	&		&		&		\\
(15,6)	&	36	&	1.49	&	156	&	16.1	&	14	&	0.15				&	1.722	&		&		&		\\
(16,4)	&	36	&	1.46	&	157	&	10.9	&	19	&	0.20				&	1.733	&		&		&		\\
(17,2)	&	36	&	1.44	&	160	&	5.5	&	22	&	0.24				&	1.738	&	1.966	&		&		\\
(18,0)	&	36	&	1.43	&	162	&	0.0	&	23	&	0.26				&	1.735	&		&		&		\\
(13,13)	&	39	&	1.79	&	130	&	30.0	&	0	&	0.00				&	1.541	&		&		&		\\
(14,11)	&	39	&	1.72	&	134	&	26.0	&	3	&	0.03				&	1.563	&		&		&		\\
(15,9)	&	39	&	1.67	&	141	&	21.8	&	7	&	0.08				&	1.615	&		&		&		\\
(16,7)	&	39	&	1.62	&	143	&	17.3	&	11	&	0.12				&	1.632	&		&		&		\\
(17,5)	&	39	&	1.59	&	146	&	12.5	&	15	&	0.16				&	1.646	&		&		&		\\
(19,1)	&	39	&	1.55	&	148	&	2.5	&	19	&	0.22				&	1.65	&		&		&		\\
(15,12)	&	42	&	1.86	&	125	&	26.3	&	3	&	0.03				&	1.487	&		&		&		\\
(16,10)	&	42	&	1.80	&	129	&	22.4	&	6	&	0.07				&	1.513	&		&		&		\\
(17,8)	&	42	&	1.76	&	131	&	18.3	&	9	&	0.10				&	1.53	&		&		&		\\
(18,6)	&	42	&	1.72	&	135	&	13.9	&	12	&	0.12				&	1.558	&		&		&		\\
(19,4)	&	42	&	1.69	&	137	&	9.4	&	14	&	0.15				&	1.562	&		&		&		\\
(20,2)	&	42	&	1.67	&	138	&	4.7	&	16	&	0.17				&	1.57	&		&		&		\\
(21,0)	&	42	&	1.67	&	139	&	0.0	&	17	&	0.20				&	1.564	&		&		&		\\
(15,15)	&	45	&	2.06	&	112	&	30.0	&	0	&	0.00				&	1.405	&		&		&		\\
(19,7)	&	45	&	1.85	&	124	&	15.1	&	10	&	0.10				&	1.474	&		&		&		\\
(20,5)	&	45	&	1.82	&	126	&	10.9	&	12	&	0.13				&	1.483	&		&		&		\\
(21,3)	&	45	&	1.80	&	128	&	6.6	&	14	&	0.15				&	1.493	&		&		&		\\
(22,1)	&	45	&	1.79	&	129	&	2.2	&	15	&	0.17				&	1.5	&	1.641	&		&		\\
(16,16)	&	48	&	2.20	&	106	&	30.0	&	0	&	0.00				&	1.326	&		&		&		\\
(17,14)	&	48	&	2.13	&	109	&	26.8	&	2	&	0.02				&	1.345	&		&		&		\\
(19,10)	&	48	&	2.03	&	114	&	19.8	&	6	&	0.07				&	1.374	&		&		&		\\
(20,8)	&	48	&	1.98	&	116	&	16.1	&	8	&	0.08				&	1.399	&		&		&		\\
(21,6)	&	48	&	1.95	&	119	&	12.2	&	10	&	0.10				&	1.401	&		&		&		\\
(22,4)	&	48	&	1.93	&	120	&	8.2	&	11	&	0.12				&	1.411	&		&		&		\\
(23,2)	&	48	&	1.91	&	121	&	4.1	&	12	&	0.14				&	1.413	&		&		&		\\
(24,0)	&	48	&	1.91	&	122	&	0.0	&	13	&	0.15				&	1.411	&	1.568	&		&		\\
(19,13)	&	51	&	2.21	&	105	&	23.8	&	3	&	0.03,	0.10$^{a}$	&		&		&		&	2.423	\\
(21,9)	&	51	&	2.12	&	107	&	17.0	&	7	&	0.07,	0.22$^{a}$	&		&		&	2.378	&		\\
(22,7)	&	51	&	2.08	&	111	&	13.4	&	8	&	0.09				&	1.345	&		&		&		\\
(23,5)	&	51	&	2.05	&	112	&	9.6	&	10	&	0.10				&	1.356	&		&		&		\\
(24,3)	&	51	&	2.03	&	114	&	5.8	&	11	&	0.11				&	1.358	&		&		&		\\
    \hline
  \end{tabular*}
\end{table*}

\begin{table*}
\small
  \caption{\ Continuation of Table \ref{table-carpet1}: Basic parameters, optical transition energies, and phonon frequencies of $\nu = 0$ (or mod 0), single-wall carbon nanotubes, produced as a vertically aligned array with a diameter range from 2.03 to 2.75~nm.   All quantities are defined as before.  RBM frequencies and $E_{ii}$ values taken from Reference \onlinecite{Doornetal08PRB}.  $^{a}$Calculated trigonal warping splitting is for $\Delta E_{22}^{M}$.}  
  \label{table-carpet2}
  \begin{tabular*}{\textwidth}{@{\extracolsep{\fill}}lllllllllll}
    \hline
   ($n,m$)	&	$2n$+$m$	&	$d_{t}$ (nm)	&	RBM (cm$^{-1}$)	&	$\alpha$ (deg)	&	$E_{g}$ (eV)$^{\dagger}$	&	$\Delta E_{ii}$ (eV)$^{*}$	&	$E_{11}^{M-}$ (eV)	&	$E_{11}^{M+}$ (eV)	&	$E_{22}^{M-}$ (eV)	&	$E_{22}^{M+}$ (eV)	\\
    \hline
(25,1)	&	51	&	2.03	&	113	&	1.9	&	11	&	0.13				&	1.359	&	1.494	&		&		\\
(23,8)	&	54	&	2.21	&	103	&	14.4	&	7	&	0.07				&	1.285	&		&		&		\\
(24,6)	&	54	&	2.18	&	105	&	10.9	&	8	&	0.08				&	1.305	&		&		&		\\
(25,4)	&	54	&	2.16	&	106	&	7.3	&	9	&	0.10				&	1.311	&		&		&		\\
(26,2)	&	54	&	2.15	&	107	&	3.7	&	10	&	0.11				&	1.317	&	1.408	&		&		\\
(27,0)	&	54	&	2.14	&	108	&	0.0	&	10	&	0.13				&	1.312	&	1.408	&		&		\\
(21,15)	&	57	&	2.49	&		&	24.5	&	2	&	0.03,	0.07$^{a}$	&		&		&		&	2.204	\\
(24,9)	&	57	&	2.35	&	99	&	15.3	&	6	&	0.06, 0.20$^{a}$	&		&		&	2.187	&		\\
(25,7)	&	57	&	2.31	&	98	&	12.0	&	7	&	0.07, 0.24$^{a}$	&		&		&	2.203	&		\\
(26,5)	&	57	&	2.29	&	101	&	8.6	&	8	&	0.08, 0.27$^{a}$	&		&	1.337	&		&	2.424	\\
(27,3)	&	57	&	2.27	&	102	&	5.2	&	9	&	0.10				&		&	1.347	&		&		\\
(28,1)	&	57	&	2.26	&	102	&	1.7	&	9	&	0.11, 0.33$^{a}$	&		&	1.359	&	2.215	&		\\
(30,0)	&	60	&	2.38	&	96	&	0.0	&	8	&			0.31$^{a}$	&		&		&	2.143	&		\\
(27,9)	&	63	&	2.58	&	90	&	13.9	&	5	&			0.18$^{a}$	&		&		&		&	2.184	\\
(33,0)	&	66	&	2.62	&	89	&	0.0	&	7	&			0.25$^{a}$	&		&		&	1.996	&	2.19	\\
(33,3)	&	69	&	2.75	&	86	&	4.3	&	6	&			0.21$^{a}$	&		&		&	1.957	&		\\
    \hline
  \end{tabular*}
\end{table*}

\subsection{Resonant Rayleigh Scattering}

Resonant Rayleigh scattering or elastic light scattering has recently become a powerful tool for the optical study and rapid characterization of nanotube samples.\cite{SfeiretAl04Science, SfeiretAl06Science, LiuetAl12NatureNano}  Applicable to individual tubes deposited or grown on a substrate over a trench, the technique can rapidly identify optical transitions due to its large scattering intensity when using a broadband, tunable excitation source such as supercontinuum-generated white light.  Sfeir \textit{et al.}\cite{SfeiretAl04Science}~made the observation of excitation spectra consistent with ``metallic" species exhibiting trigonal warping splitting.  However, the individual tubes measured were not ($n,m$)-identified.  In a subsequent study by Sfeir \textit{et al.},\cite{SfeiretAl06Science} where individual tubes were ($n,m$)-identified using electron beam diffraction, measurements on metallic species revealed a single optical transition feature for armchair species and a two-peak feature for non-armchair, $\nu$~=~0 species due to trigonal warping.  Most recently, using the approach of Sfeir \textit{et al.}, Liu \textit{et al.}\cite{LiuetAl12NatureNano} combined electron beam diffraction with \textit{in-situ} Rayleigh scattering to measure 206 optical transitions of ($n,m$)-identified nanotubes.  In particular, for $\nu$~=~0 species, optical transitions were measured ranging from $E_{11}^{M}$ to $E_{33}^{M}$, observing trigonal warping splitting for non-armchair tube.  From this broad range of data, covering nanotubes as small in diameter as (10,10) to as large as (31,31), an empirical equation was produced to predict optical transition energies with appropriate ($n,m$)- and $E_{ii}$-dependent parameters.  However, it should be noted that small-diameter metals with optical transitions $>$2.5~eV were not studied, which is where most deviations from graphene-like behavior occur (small-diameter metals are the focus of our work presented later). 







\section{Methods for Enrichment of Armchair Carbon Nanotubes}
\label{Separation}
Optical investigations of armchair and other $\nu$~=~0 nanotubes have been often impaired by multiple issues related to sample quality and signal detection.  Due to an inability at this time for researchers to exactly control and tune the chirality distribution of species grown in nanotube syntheses, as-produced nanotube materials are plagued with issues of sample heterogeneity with respect to physical parameters such as nanotube length, diameter, chiral angle, electronic type, and aggregation state.  Optical measurements on metallic SWCNTs contained within macroscopic ensembles composed of heterogeneous material, such as colloidal suspensions, powders, and thin films, typically suffer from problems such as spectral overlap wherein the response of a particular optical mode of a particular ($n,m$) species is obscured by overlap with the optical response from other nanotube species.  Furthermore, even when spectral overlap is not an issue and spectral features are well-defined, optical spectra containing contributions from both metallic and semiconducting species can be dominated by signal from semiconductor components due to their often larger optical response.  This is particularly true for resonant Raman scattering.  

One approach to circumvent these sample heterogeneity issues is in the use of single-nanotube micro-spectroscopy where a single nanotube is deposited or grown on a substrate or host medium and its optical response is subsequently measured.  While this allows for clear measurement of the optical properties of a particular SWCNT species, such measurements typically produce small signal and still suffer from the problem of lack of chirality control (i.e., one does not have fine control of what species can be studied since the growth process cannot be controlled).  This results in the need to do many measurements on many different tubes to find suitable metallic tubes for study, which can be time-consuming.  Additionally, because of the one-by-one nature of single-nanotube spectroscopy, studies are not necessarily statistically representative of a particular nanotube material.  

Another approach in the optical study of armchair SWCNTs that avoids sample issues while simultaneously avoiding the experimental difficulties of single-nanotube micro-spectroscopies is to produce macroscopic ensembles of SWCNTs enriched in armchair nanotube species by post-synthesis separation methods.   Researchers have attempted such post-production separations since the early days of nanotube research via methods such as selective chemical functionalization,\cite{StranoetAl03Science} complexation,\cite{ChenetAl03NL} dielectrophoresis,\cite{KrupkeetAl03Science} etc.  However, these methods often modify the properties of the nanotubes or do not produce sufficient material for ensemble optical measurements.  Additionally, the specificity of separation to one specific chirality of such methods is often low. 

Recently, researchers have appealed to techniques normally reserved for biochemistry to separate SWCNTs due to the large volume of work demonstrating high-purity and specificity in separation of biological macromolecules, such as DNA and proteins, based upon molecular mass and chemical moiety.  Given the similarity of carbon nanotubes in physical dimensions and mass to DNA and proteins, such an approach is logical and fruitful.  In particular, bio-inspired methods for separation such as centrifugation and chromatography have been explored with great success to separate SWCNTs based on various geometrical and electronic parameters.  Here, we discuss the two most successful methods, density gradient ultracentrifugation and DNA-based ion-exchange chromatography, both of which lead to the production of armchair-enriched macroscopic ensembles suitable for optical measurements.

\subsection{Density Gradient Ultracentrifugation}
\label{DGU}

\begin{figure}
\centering
\includegraphics[scale=.35]{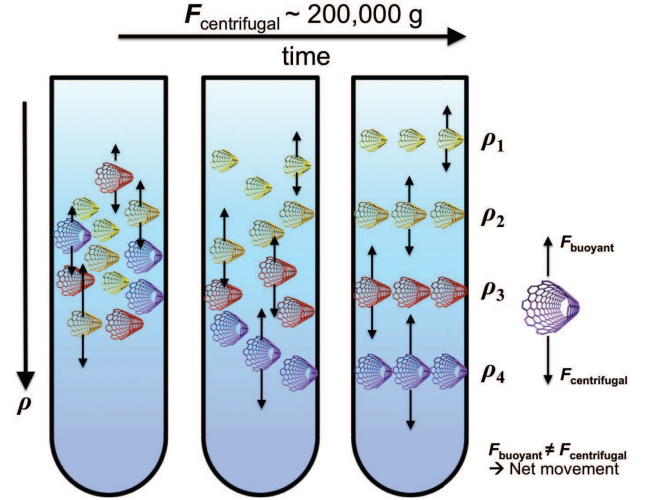}
\caption{Illustration of the density gradient ultracentrifugation method as applied to the separation of SWCNTs.  SWCNTs of different mass density (represented by different color tubes) are loaded into a centrifuge tube containing a mass density gradient where a strong centrifugal field is applied.  Due to a competition between the applied centrifugal force and the resultant opposing buoyant force, SWCNTs migrate through the gradient in space until equilibrium is reached and the forces are balanced.  At this point, the SWCNTs have arrived in a region in the gradient possessing their respective mass density.}
\label{fig-DGUsep4}
\end{figure}

Density gradient ultracentrifugation (DGU) separates different molecular species by exploiting differences in their respective mass densities as they travel through a viscous medium under a large applied centrifugal field.   Briefly, an aqueous suspension of surfactant-suspended SWCNTs travel through a mass density gradient via ultracentrifugation ($\sim$200,000g for several hours).  Due to the variation in density of the gradient, different density nanotube species will migrate towards different regions of the gradient, under the applied force of the centrifuge versus the buoyant force of each nanotube due to its mass density, until an equilibrium is reached.  At this point, the nanotubes are now sufficiently separated in vertical distance in the gradient so that they may be extracted using simple fractionation techniques (see diagram in Fig.~\ref{fig-DGUsep4}).   As the physical species being separated is an individual SWCNT surrounded by a surfactant micelle and several layers of hydration, the intrinsic differences in mass density of different chirality nanotubes (and hence the chiral-specific buoyant forces) are primarily due to the varying dimensions of different chiralities (diameter and length) as well as the size and surface coverage of the surfactant micelle around the nanotube, which can be a complex function of nanotube diameter, surfactant type, ionic concentration, and sterics.  However, this complexity can also lead to high specificity of separation as well.  

DGU was first applied to SWCNTs by Arnold \textit{et al.}\cite{Arnoldetal05NL} using aqueous suspensions of DNA-wrapped HiPco- and CoMoCAT-produced SWCNT materials.  There, they demonstrated the efficacy of DGU for separating SWCNTs by observing diameter-based sorting as measured by optical absorption spectroscopy.  Arnold \textit{et al.}~subsequently demonstrated DGU using anionic surfactants commonly used to suspend SWCNTs by ultrasonication and centrifugation, namely sodium dodecyl sulfate (SDS) and sodium cholate (SC).\cite{ArnoldetAl06NatureNano}  In performing DGU with the single surfactant SC, it was shown that not only could diameter-sorting be achieved with greater efficiency than their previous DNA-suspended work,\cite{Arnoldetal05NL} but that with multiple iterations of DGU, single-chirality semiconducting samples, namely (6,5) and (7,5), could be produced with CoMoCAT SG65 grade nanotube material, as demonstrated by optical absorption and photoluminescence excitation (PLE) spectroscopy measurements.  In the same work, it was shown that the addition of a second surfactant, SDS, to the density gradient, resulted in sorting by electronic type.  Specifically, laser-ablation-produced SWCNT material could be separated into thin, concentrated fractions of nearly pure metallic and semiconducting nanotubes as shown by optical absorption.  This opened the door for many different types of measurements on metallic nanotubes, ranging from DC electrical transport to optical spectroscopy.  To confirm their type separation, thin films of purely metallic and semiconducting nanotubes were fabricated by vacuum filtration.  Significantly lower sheet resistivity was observed for the metal-enriched film as opposed to the semiconducting film, as would be expected.\cite{ArnoldetAl06NatureNano}  

On the heels of this breakthrough in SWCNT separation science utilizing commonly employed surfactants as opposed to DNA, a flood of research followed, improving and expanding the DGU technique further.  Yanagi \textit{et al.}\cite{YanagietAl08APE} added a third surfactant to the combination of SDS and SC in their electronic-type sorting with the addition of sodium deoxycholate (DOC), resulting in an expansion in vertical separation distance of the thin type-separated fractions observed by Arnold \textit{et al.}.  This resulted in significantly more facile fractionation of the separated products, which they demonstrated on CoMoCAT, HiPco, and laser ablation carbon nanotube material.  In addition to optical absorption and sheet resistance measurements on their metal-enriched material, resonant Raman measurements were taken in the RBM and G-band frequency regions to further confirm enrichment.  Green \textit{et al.}\cite{GreenHersam08NL} similarly expanded the work of Arnold \textit{et al.},\cite{ArnoldetAl06NatureNano} using the SDS-SC surfactant combination to separate CoMoCat, HiPco, and arc-discharge SWCNT materials, forming semitransparent, conductive thin films from the metal-enriched fractions and measuring optical transparency and sheet resistance of said films for possible applications to flexible electronics.  Taking a slightly different approach, Niyogi \textit{et al.}\cite{NiyogietAl09JACS} used a single surfactant, SDS, along with the addition of electrolytes, such as alkali and alkaline earth salts, to tune the interfacial nanotube-micelle interaction to achieve bulk type enrichment of metals and semiconductors in HiPco material, significantly increasing the scale of enriched product produced.  Enrichment was confirmed by a combination of optical absorption, PLE, and Raman measurements.  Further improving upon their single-chirality work, Green \textit{et al.}\cite{GreenetAl09NanoRes} demonstrated that single-chirality (6,5) material could be further purified into left- and right-handed optical enantiomers using the chiral surfactant SC.  This separation is possible due to a preference of the rigid and planar SC molecule for the left-handed (6,5) enantiomer over the right-handed (6,5), resulting in a difference in surfactant coverage of the nanotube and hence mass density difference between enantiomers.  Extensions of the DGU method have allowed separation of double-wall carbon nanotubes (DWCNTs) from SWCNTs based on diameter,\cite{GreenHersam09NatureNano} DWCNTs by outer-wall electronic type,\cite{HuhetAl10JPCC, GreenHersam11ACS} SWCNTs by nanotube length,\cite{FaganetAl08Langmuir} and even SWCNTs by water-filling (empty versus water-filled, i.e., end-capped versus open-ended).\cite{CambreetAl11ACIE, FaganetAl11ACS}

\begin{figure}
\centering
\includegraphics[scale=1.04]{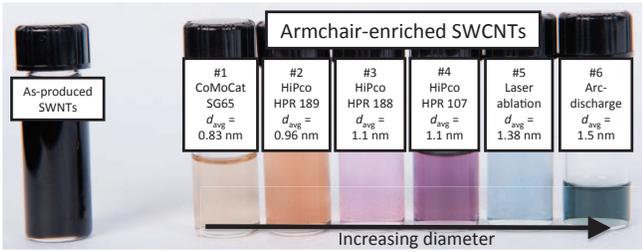}
\caption{Pictures of armchair-enriched SWCNT suspensions. The black, ``as-produced" vial on the left is typical of unsorted, SWCNT materials. On the right, various armchair-enriched samples with different diameters exhibit different, distinct colors. Reproduced from Reference \onlinecite{HarozetAl12JACS}.}
\label{fig-colorspic}
\end{figure}

Motivated by these advancements in separation science and also the lack of specific ($n,m$) composition information in metal-enriched materials, we focused our approach to produce metal-enriched SWCNT material that could be identified as a function of ($n,m$) composition.  We employed a three-surfactant DGU method similar to that of Yanagi \textit{et al.}\cite{YanagietAl08APE} and described in detail in H\'{a}roz \textit{et al.}\cite{HarozetAl10ACS}~for HiPco SWCNTs (batch 188.2).
For SWCNT material produced by other synthesis methods such as CoMoCAT, other HiPco batches, laser ablation, and arc-discharge material, similar gradient parameters were used as described in H\'{a}roz \textit{et al.}\cite{HarozetAl12JACS}  Figure \ref{fig-colorspic} shows a photograph of typical, unsorted SWCNT material appearing black on the left side and the resulting metal-enriched layers from CoMoCAT, three different HiPco batches, laser ablation, and arc-discharge materials appearing yellow, orange, pink, purple, cyan, and green, respectively, on the right side.  As will be discussed in later sections, resonant Raman measurements revealed not only that the topmost fraction is highest in metallic species content but also that the chiral distribution of $\nu$~=~0 species has been affected with preferential enrichment towards armchair chiralities.  

Taking the results of the aforementioned DGU studies together,\cite{ArnoldetAl06NatureNano, YanagietAl08APE, GreenHersam08NL, NiyogietAl09JACS, HarozetAl10ACS, HarozetAl12JACS} some generalizations regarding surfactant-based DGU can be made.  First of all, the SDS surfactant is common and essential to all anionic-surfactant-based DGU methods that attempt separation by electronic type.  This is most likely due to an electronic-type-specific binding affinity difference between SDS and metallic/semiconducting nanotubes.  However, SDS alone cannot achieve type separation as demonstrated by experiments performed by Arnold \textit{et al.}\cite{ArnoldetAl06NatureNano} 
An additional component must be added to further differentiate metal and semiconductor mass densities.  In the case of two-surfactant systems, SC is employed to increase soluble SWCNT concentration, resulting in thin, concentrated, closely spaced bands corresponding to semiconductor- and metal-enriched fractions.  In the case of SDS-electrolyte systems, the addition of alkali salts changes the micelle dimensions in a type-specific manner, resulting in mass density changes that produce broader bands ranging from highly metallic to highly semiconducting fractions.  In the case of three-surfactant systems like those used here, involving SDS, SC, and DOC, one would expect similar results to that of the two-surfactant system due to the almost identical molecular structures of SC and DOC.  However, two important differences exist, which may result in the significantly broader bands observed.  One is the significantly higher buoyancy of DOC as compared to SC.  This causes the normally thin metal band observed in two-surfactant DGU to become expanded in vertical separation distance, allowing higher resolution of the metallic fraction and finer fractionation.  Secondly, as first suggested by Green \textit{et al.},\cite{GreenetAl09NanoRes} the cholate family of chiral surfactants (SC and DOC) can interact selectively with certain preferred geometries of nanotubes (namely, armchair and other large chiral angle species).  In the case of DOC, this preferred steric interaction coupled with its higher buoyancy may be responsible for the simultaneous observation of highest metal enrichment with highest armchair enrichment in the topmost metallic fractions.  As one extracts lower-lying, higher density fractions, both bulk metal and armchair purity decrease.  In Section \ref{RRS-RBM}, it is shown that this chiral-angle-specific steric interaction between SWCNTs and DOC can be observed even in unsorted nanotube material suspended by DOC.



\subsection{DNA-based Separation}
\label{DNA}

\begin{figure}
\centering
\includegraphics[scale=.55]{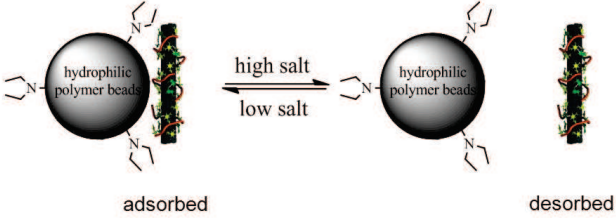}
\caption{Anion-exchange chromatography separation of DNA-SWCNTs. Anion exchange columns are packed with micro-sized polymer or silica beads that are surface-grafted with positively charged functional groups. DNA-SWCNTs injected into the column can be either adsorbed on the beads or remain free, depending on the concentration and chemical identity of the salt in the mobile phase. The equilibrium is also dependent on SWCNT structure, leading to fractionation of DNA-SWCNTs when they are eluted from the column by a salt gradient.  By tailoring specific ssDNA strands to form ordered structures only on a specific ($n,m$) SWCNT surface, preferential adsorption and retention can be achieved leading to single-chirality samples.} 
\label{fig-DNAsep2}
\end{figure}

While DGU can produce ensembles strongly enriched in several armchair species, ensembles of a single-chirality armchair species have remained elusive, which for many experiments represents the most ideal sample for understanding the physics of one-dimensional metals.  As an alternative to DGU and continuing forward with the idea of chirality-specific interactions to achieve such a goal, DNA-based ion-exchange chromatography (IEX) has been used to purify bulk, unsorted nanotube material into \textit{single-chirality} armchair carbon nanotubes.  For a recent review of the approach, see reference~\cite{TuZheng08NanoRes}. IEX separates SWCNTs based on differential adsorption and desorption of DNA-wrapped SWCNTs (DNA-SWCNTs) on chemically functionalized resins packed in an IEX column (Fig.~\ref{fig-DNAsep2}).\cite{TuZheng08NanoRes}  The choice of single-stranded DNA (ssDNA) sequence for SWCNT wrapping plays a key role in the separation process: the wrapping structure of ssDNA may be ordered or disordered depending on the ssDNA sequence and the SWCNT chirality, resulting in differential adsorption and retention of different types of SWCNTs when they are eluted by a salt gradient.\cite{TuZheng08NanoRes, ZhengetAl03NatMat, ZhengetAl03Science, TuetAl09Nature, TuetAl11JACS}  Initially, an electrostatic interaction-based separation mechanism was proposed for the low resolution electronic-type and diameter separation observed in earlier work.\cite{ZhengetAl03NatMat, ZhengetAl03Science}  However, observations from purification of single-chirality nanotubes had led to the proposal that other factors, such as hydrophobic and van der Waals interactions between DNA-SWCNTs and IEX resin may be more important in the IEX separation of single-chirality SWCNTs when short ssDNA sequences are used.\cite{TuZheng08NanoRes, TuetAl09Nature, TuetAl11JACS}

To identify proper ssDNA sequences for single-chirality SWCNT purification,  a search strategy was devised to survey the vast ssDNA library for specific sequences that enable IEX purification of particular ($n,m$) nanotubes from a synthetic mixture (HiPco).\cite{TuetAl09Nature}  The survey was designed to span the maximum sequence space with a minimum number of test sequences via the use of simple nucleotide repeats.  Out of $>$~350 sequences tested, $\sim$~20 semiconducting tube recognition sequences in the 9-14-mer length range were identified.   These short ssDNA sequences, capable of enriching 12 semiconducting SWCNTs from HiPco, were found to conform to the pattern of single purine/multiple pyrimidine repeats, such as TTA and TTTA.\cite{TuetAl09Nature}  The pattern seems to allow ssDNA to form an ordered structure on the surface of a particular type of SWCNT through inter-strand hydrogen bonding,\cite{RoxburyetAl12NL} minimizing hydrophobic and van der Waals interactions between SWCNTs and the column resin, and resulting in early elution of the SWCNTs.  

Since the single purine/multiple pyrimidine repeats produced the enrichment of semiconducting nanotubes, it is reasonable to expect that recognition sequences for metallic tubes may be found outside the sequence space defined by such a pattern.  In view of the structural similarities between metallic and semiconducting tubes, an evolutionary approach was taken to find metallic tube recognition sequences by limiting the candidate pool only to those that are direct descendants of semiconducting tube recognition sequences.  Such an approach was expected to balance the need to deviate from the purine/pyrimidine pattern and the desire to conserve sequence features that are believed to sustain a 2D DNA sheet structure for ordered SWCNT wrapping.  Such an evolutionary approach is analogous in spirit to the well-documented genetic algorithm for solving a wide variety of optimization problems. 
 In one set of searches, a new sequence pattern, i.e., AATT repeats, was selected for testing based on previously identified ATT and ATTT repeats for semiconducting tube purification.  This selection was rationalized by a proposed structure model that invokes interchain hydrogen bonding interactions to form an ordered DNA wrapping sheet for SWCNTs.  Similar to the ATT and ATTT motif, the AATT repeats can form A:T:A:T hydrogen-bond quartets and conserve the 2D sheet feature of the ATT and ATTT repeats.  Through an exhaustive test of the AATT pattern in the 9-14-mers range, a sequence ATTAATTAATTAAT was found to allow purification of (6,6) armchair tubes.   For (7,7) separation, a single-point mutation on previously identified sequences was carried out with the limitation that only single-base purine-to-purine or pyrimidine-to-pyrimidine replacement was allowed.  By replacing one T at a time by C, TTATTATTATTATT, a DNA sequence originally assigned to (8,3) purification, led to the identification of TTATTACTATTATT for (7,7) purification.  

While both DGU and DNA-based separation methods described here can produce highly enriched armchair SWCNT aqueous suspensions, it should be noted that only sub-microgram to as much as single milligram quantities of armchair SWCNTs can be produced and usually only after combining multiple applications of either separation technique.  This is a consequence not of the chirality composition of the initial parent SWCNT material before separation but rather the overall efficiency of the separation technique.  Although these techniques are not ideal for large-scale separation of gram and kilogram quantities of as-produced SWCNT material, they produce more than sufficient amounts for spectroscopic interrogation.

\section{Optical Absorption and Colors of Armchair Carbon Nanotubes}
\label{absorption}




\begin{figure}
\centering
\includegraphics[scale=0.53]{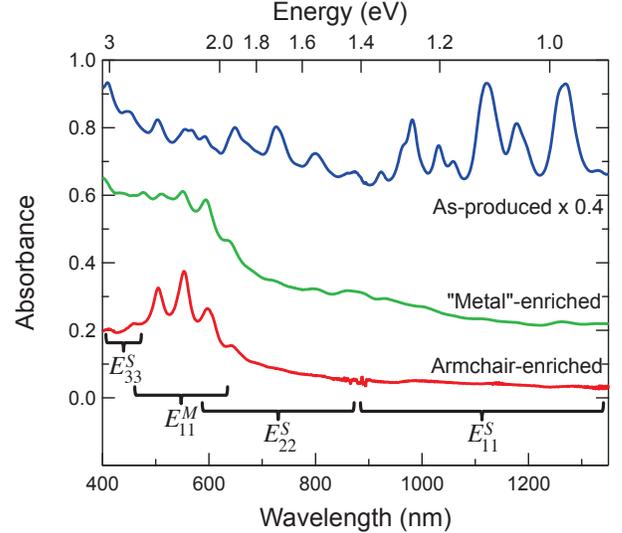}
\caption{Optical absorption of three HiPco nanotube suspensions studied: unsorted (AP-SWCNT, blue, top), ``metal''-enriched (ME-SWCNT, green, middle), and armchair-enriched (AE-SWCNT, red, bottom).  Note ME-SWCNT and AE-SWCNT show significant suppression of features due to $\nu$~=~$\pm$1 SWCNTs as compared to AP-SWCNT, indicating the elimination of $\nu$~=~$\pm$1 species and hence $\nu$~=~0 enrichment by the DGU process.  Reproduced from Reference \onlinecite{HarozetAl11PRB}.}
\label{fig-mixedAbs}
\end{figure}
Optical absorption spectroscopy is one of the most fundamental and, in many cases, experimentally simple optical measurements that can made on a nanomaterial.  The optical absorption process itself is easy to understand in part because of its relationship with the optical property of ``apparent color" of materials in the physical world.  In the case of SWCNTs, optical absorption spectroscopy can provide fundamental quantities, such as optical transition energies, transition oscillator strengths, and transition line widths and line shape.  Then with the optical features assigned to known ($n,m$) chiralities, other physical properties such as diameter and chiral distribution, ($n,m$) concentration, and extinction coefficient can be determined.  For enriched samples produced by methods such as DGU and DNA-based chromatography, optical absorption can provide information on separation purity and efficiency.  However, this is possible only when optical features can be clearly identified by chirality.

Figure \ref{fig-mixedAbs} shows optical absorption spectra of three HiPco SWCNT samples.  The unsorted HiPco sample (AP-SWCNT, Fig.~\ref{fig-mixedAbs} top, blue trace) displays all the optical absorption features typically observed in surfactant-suspended dispersions\cite{OConnelletAl02Science} with peaks corresponding to the first ($E_{11}^{\rm S}$, 850-1600~nm) and second ($E_{22}^{\rm S}$, 570-850~nm) interband transitions of $\nu$~=~$\pm$1 tubes and the first ($E_{11}^{\rm M}$, 440-650~nm) interband transitions of $\nu$~=~0 tubes.  Based on absorption peak area estimates, this sample contains $\sim$40\% $\nu$~=~0 nanotubes,\cite{HarozetAl10ACS} which agrees reasonably well with the statistically expected value of 34\% from the number of all possible ($n$,$m$) species contained within the diameter range (0.6-1.4~nm) of this particular material, assuming that all species are equally probable.  This type of estimate must be taken cautiously, however, due to the overlap of spectral excitation regions corresponding to semiconducting and metallic nanotubes, resulting in uncertainty as to appropriate regions to integrate over for the respective contributions.  Furthermore, an ($n,m$)-specific interpretation of the spectrum is not possible due to the overlap of optical transitions of several species, resulting in observed absorption features that may be due to multiple chiralities.  After DGU, however, a significant suppression of $\nu$~=~$\pm$1 features (650-1350~nm) is observed in both the ``metal"-enriched (ME-SWCNT, Fig.~\ref{fig-mixedAbs} middle, green trace) and armchair-enriched (AE-SWCNT, Fig.~\ref{fig-mixedAbs} bottom, red trace) samples with $\nu$~=~0 purity estimates around 90\% and 98\%, respectively.  A clear delineation of the extent of the region containing metallic species is visible.  

While the overlap between semiconductors and metals has now been removed, individual absorption peaks are still not well-defined in the ME-SWCNT sample due to the overlap of optical transitions from members of a metallic $2n$+$m$ family for each absorption peak.  This can be understood from a quick examination of the Kataura plot in the diameter regime of HiPco SWCNTs.  For members of the same $2n$+$m$ family, for example, Family 27, ($n,m$) species are not separated by more than 60~meV in their $E_{11}^{M-}$ transition energies, which is comparable to the line width of the absorption peak due to a single $\nu$~=~0 species.  Additionally, due to trigonal warping, one would expect contributions from the $E_{11}^{M+}$ transitions, as recently predicted by Malic \textit{et al.}\cite{MalicetAl10PRB}~The AE-SWCNT shows similar characteristics to the ME-SWCNT sample with regards to semiconductor removal.  However, the absorption features are better defined, exhibiting sharper peaks with an increased peak-to-valley ratio.  While both DGU samples are strongly enriched in $\nu$~=~0 tubes, Raman studies based on radial breathing mode (RBM) intensities show that ME-SWCNT samples are a {\em bulk} enrichment of all $\nu$~=~0 species\cite{NiyogietAl09JACS} whereas AE-SWCNT samples are chiral-angle-selective toward armchair ($n$ = $m$) chiralities.\cite{HarozetAl10ACS}  This accounts for the sharper absorption features observed in the AE-SWCNT sample due to the sizable reduction of the number of overlapping peaks from the various $\nu$~=~0 species and $E_{11}^{M}$ branches.  

\begin{figure}
\centering
\includegraphics[scale=0.5]{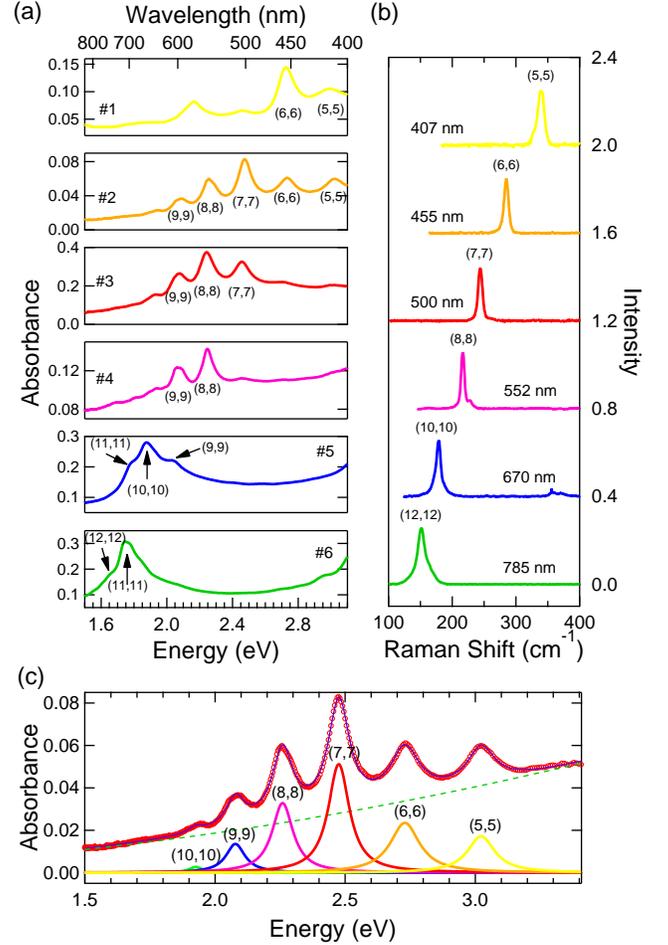}
\caption{(a) Optical absorption spectra of armchair-enriched SWCNT suspensions, corresponding to the samples identified in Fig.~\ref{fig-colorspic}.  The main absorption features move toward higher energy (shorter wavelength) with decreasing SWCNT diameter.  (b) Associated single-line resonant Raman radial breathing mode spectra for each enriched sample, further confirming the armchair enrichment of these samples and the peak assignments in absorption.  (c) Resulting fit (thin purple line) of the optical absorption spectrum of armchair-enriched sample \#2 (HiPco HPR 189.2, red open circles) to a sum of six Lorentzian peaks (thick, multi-color peaks), one for each armchair (10,10) through (5,5), on top of an absorption baseline (dashed green line).  Reproduced from Reference \onlinecite{HarozetAl12JACS}.}
\label{fig-AbsErikJACS}
\end{figure}

To examine the optical absorption properties of armchair SWCNTs more closely, H\'{a}roz \textit{et al.}\cite{HarozetAl12JACS}~produced armchair-enriched suspensions from SWCNTs produced by various synthesis methods that differed in diameter distribution.  Briefly, armchair enrichment via DGU was performed on parent nanotube material produced by the CoMoCAT (sample \#1), HiPco (samples \#2, 3 and 4), laser ablation (sample \#5), and arc-discharge (sample \#6) methods.  As can be seen from Fig.~\ref{fig-colorspic}, each enriched material exhibits a distinct color with corresponding unique absorption structure shown in Fig.~\ref{fig-AbsErikJACS}a.  Each absorption peak is attributed to predominantly one armchair species as indicated in the spectra, supported by resonant Raman RBM spectra shown in Fig.~\ref{fig-AbsErikJACS}b.  A clear trend of the peak position of each absorption feature can be observed with optical transition energy increasing with decreasing diameter (arc-discharge to CoMoCAT).  This ``tunability" shows an ability to control optical absorption in the visible region of a metallic nanotube sample by careful selection of the starting SWCNT material used for DGU.  Additionally, it demonstrates the generality and applicability of the DGU technique to most readily available SWCNT materials.  Based on the combination of the spectral simplicity as well as the supporting Raman spectra, more insightful information can be extracted from the absorption spectra through a fitting analysis.  Using one of the HiPco samples as an example (sample \#2), the spectrum was fitted with a sum of Lorentzian peaks, one peak for each observed absorption feature corresponding to armchair species (10,10) through (5,5), on top of a baseline fitted to a polynomial, as shown in Fig.~\ref{fig-AbsErikJACS}c.  Excellent agreement between the experimental data and simulated spectrum is observed, lending to the validity of our approach.  

Physically, the use of highly symmetric Lorentzian features in peak fitting and the resulting good agreement with measured spectra indicate a significant contribution of excitons to the optical properties of armchair SWCNTs.  If excitons were not a major contributor, one would expect a highly asymmetric absorption feature with a sharp low-energy side peaking at the optical transition energy, followed by a slowly decaying high-energy tail reflective of the $E^{-1/2}$ dependence on energy of the one-dimensional van Hove singularities in the electronic density-of-states.  Instead, one-dimensionality reduces the ability of free carriers to screen the Coulomb interaction between electrons and holes, resulting in the formation of stable, bound excitons.\cite{OgawaTakagahara91PRBRC, OgawaTakagahara91PRB}  This is supported by recent theoretical studies\cite{DeslippeetAl07NL, MalicetAl10PRB} as well as single-tube absorption measurements on a (21,21) tube.\cite{WangetAl07PRL}  Furthermore, this interaction suppresses the contribution from the continuum of states, resulting in a Sommerfeld factor, the absorption intensity ratio of an unbound exciton to the free electron-hole pair above the band edge, less than unity.\cite{OgawaTakagahara91PRBRC, OgawaTakagahara91PRB}  Armed with this knowledge, the observed colors of the different armchair-enriched suspensions in Fig.~\ref{fig-colorspic} can be explained as a consequence of a unique combination of electronic band structure and optical selection rules for armchair species.  As mentioned earlier, armchair chiralities are the only truly gapless species of SWCNTs possessing both a combination of linear bands, which cross only at the K-points of the graphene Brillouin zone, and parabolic bands.  Optical transitions between linear bands are symmetry-forbidden (see Fig.~\ref{fig-MetalBand}), causing the lowest-allowed, optical transition for polarization parallel to the tube axis to be between corresponding parabolic bands of the first valence and conduction bands.  

In other metallic nanomaterials, the observed color is controlled by their plasmon resonances, which for longitudinal plasma oscillations in SWCNTs occurs in the far-infrared region.\cite{AkimaetAl06AM}~In the case of semiconducting nanomaterials, their observed color is due to their size-dependent band gaps, which for armchair SWCNTs, is zero.  Additionally, as shown by H\'{a}roz \textit{et al.}\cite{HarozetAl12JACS} for armchair-enriched SWCNT material and by Green and Hersam\cite{GreenHersam08NL} for bulk $\nu$~=~0-enriched SWCNT material, optical absorption spectra and hence color is not affected by aggregation size.  Rather for armchair carbon nanotubes, their color is due to the sharp, narrow absorption of energy through the lowest excitonic optical transition (which is not the HOMO-LUMO separation).  This unique combination of optical and electrical properties makes armchair SWCNTs a novel, hybrid class of optical nanomaterial, apart from conventional semiconductors and metals, with optical properties like that of GaAs and electrical properties like that of copper.


\begin{figure}
\centering
\includegraphics[scale=1]{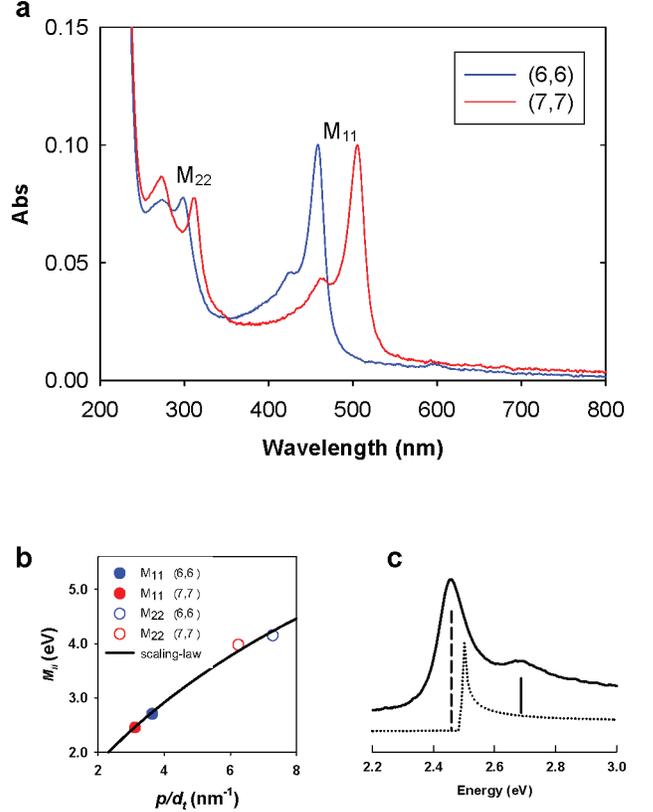}
\caption{(a) Optical absorption spectra of purified DNA-wrapped (6,6) (blue) and (7,7) (red) armchair carbon nanotubes in water. The spectral intensities were rescaled for easy comparison, but no other processing steps (e.g., baseline subtraction) were done. (b) $E^{M}_{ii}$ ($M_{ii}$) values as a function of $p/d_{t}$, where $p$~=~3 and 6 for $E^{M}_{11}$ ($M_{11}$) and $E^{M}_{22}$ ($M_{22}$), respectively, and $d_{t}$ is the tube diameter in nm. The black solid curve is a fit to the scaling-law equation given in reference~\cite{TuetAl11JACS}. (c) Line shape analysis of the $M_{11}$ peak for (7,7) nanotubes (solid trace), which can be viewed as a superposition of three transitions: discrete (vertical dashed line) and continuum (dotted trace) exciton transitions and a sideband (vertical solid line) located 0.217 eV above the discrete exciton peak.  Reproduced from Reference \onlinecite{TuetAl11JACS}.}
\label{fig-AbsMing}
\end{figure}

To probe further into the intrinsic optical structure of armchair nanotubes, further refinements in sample preparation are needed.  The single-chirality armchair sample as prepared via DNA-based IEX\cite{TuetAl11JACS} provides the ability to probe a macroscopic ensemble of a homogeneous population of armchair SWCNTs.  As shown in Fig.~\ref{fig-AbsMing}a, for both the (6,6) and (7,7) single-chirality samples, the ultimate in spectral simplicity is achieved with spectral features reflecting the presence of only one armchair species without any contamination from other nanotube species, metallic or semiconducting, as supported by resonant Raman spectra of both RBM and G-band features.  However, despite their chiral simplicity, a spectral richness is revealed in these optical absorption spectra, not previously observed.  First, absorption spectra reveal the appearance of two clusters of absorption features for each single-chirality sample.  Each cluster is made up of one strong, sharp feature and a weaker, higher-energy feature located 211 and 217~meV above the lower-energy feature, respectively, for the (6,6) and (7,7) samples.  The strong, sharp features located at 458 and 299~nm for (6,6) and 505 and 312~nm for (7,7) are attributed to $E_{11}^{M}$ and $E_{22}^{M}$ excitonic transitions, respectively, in good agreement with $E_{11}^{M}$ values measured for DNA-suspended SWCNTs.\cite{FantinietAl07CPL}  This is also supported by good agreement with a fit (shown in Fig.~\ref{fig-AbsMing}b) of the optical transition energies to an energy scaling relation, plotted as a function of $p/d_{t}$, where $p/d_{t}$ is the ratio of the carbon nanotube transition index to nanotube diameter.   Plotting in this way expresses the optical transition energy as a diameter-dependent term accounting for the quantum confinement on the graphene band structure imposed by the one-dimensional structure of a nanotube plus a logarithmic correction accounting for many-body interactions.  The weaker, higher-energy feature alongside each excitonic transition has been attributed to a phonon sideband arising from the longitudinal optical phonon.  This is supported by recent theoretical work by Bobkin \textit{et al.}\cite{BobkinetAl12PRB} and earlier less definitive experimental claims.\cite{WangetAl07PRL, ZengetAl09PRL}  

Finally, with these facts in hand, the true armchair absorption spectrum line shape can be investigated.   As shown in Fig.~\ref{fig-AbsMing}c, the absorption spectrum can be attributed to three components: 1) a strong, Lorentzian component centered at the optical transition energy and due to the discrete excitonic transition; 2) a slightly higher-energy, yet much weaker band-to-band transition continuum (weak due to Sommerfeld factor being less than 1) at $\sim$~100~meV above the discrete transition, due to exciton binding energy; and 3) and a small discrete feature due to the newly observed phonon sideband, $\sim$~200~meV above $E_{ii}^{M}$.  The reported $E_{22}^{M}$ values constitute the first measurement of second optical transitions for such small-diameter metallic nanotubes, only made possible by the chiral simplicity of the sample.  Likewise, the weaker, high-energy features have only now been observed clearly due to the chiral simplicity of single-chirality armchair samples.  Although such features should exist in armchair-enriched suspensions produced via DGU, the coincidental overlap of these features with the main optical transitions of the next smallest armchair species (next highest energy absorption feature), obscured their detection.  This highlights the importance of sample purity in probing and discovering new photophysics in carbon nanotubes to allow illumination of a more complete picture of their internal structure.    

\section{Resonant Raman Scattering Spectroscopy of Armchair-Enriched Ensemble Samples of Carbon Nanotubes}
\label{RRS-RBM}

As mentioned in Section \ref{Separation} and shown in Section \ref{absorption}, optical absorption spectra of metal-enriched fractions produced by H\'{a}roz \textit{et al.}\cite{HarozetAl10ACS} show a strong enhancement (suppression) in optical absorbance of spectral features corresponding to metallic (semiconducting) species of HiPco material relative to unsorted material, indicating that metal enrichment has occurred.  However, because of the overlap and spectral ``clustering" of optical transition energies for members of the same $2n$+$m$ family, absorption features in both the unsorted and metal-enriched samples are superpositions of multiple species and as such limit the chirality-specific information absorption spectroscopy can provide outside of bulk type enrichment.  To further investigate whether the ($n,m$) distribution has been affected by sorting methods such as DGU, resonant Raman scattering (RRS) can be used to determine ($n,m$) composition of unsorted and metal-enriched fractions of nanotube material through the use of the diameter-dependent radial breathing mode (RBM) frequency of a carbon nanotube coupled with the resonant nature of nanotube Raman scattering.  By using RRS combined with a tunable cw excitation source, Raman spectra of the RBM can be taken continuously as a function of excitation energy to map out the presence of every nanotube species contained within a given sample.  Given the appearance of the RBM regardless of an ($n,m$) species' electronic type, environment, doping or aggregation state (such factors mainly affect the optical transition energy), RRS is ideally suited for determining chiral composition.  


\begin{figure*}
\centering
\includegraphics[scale=.6]{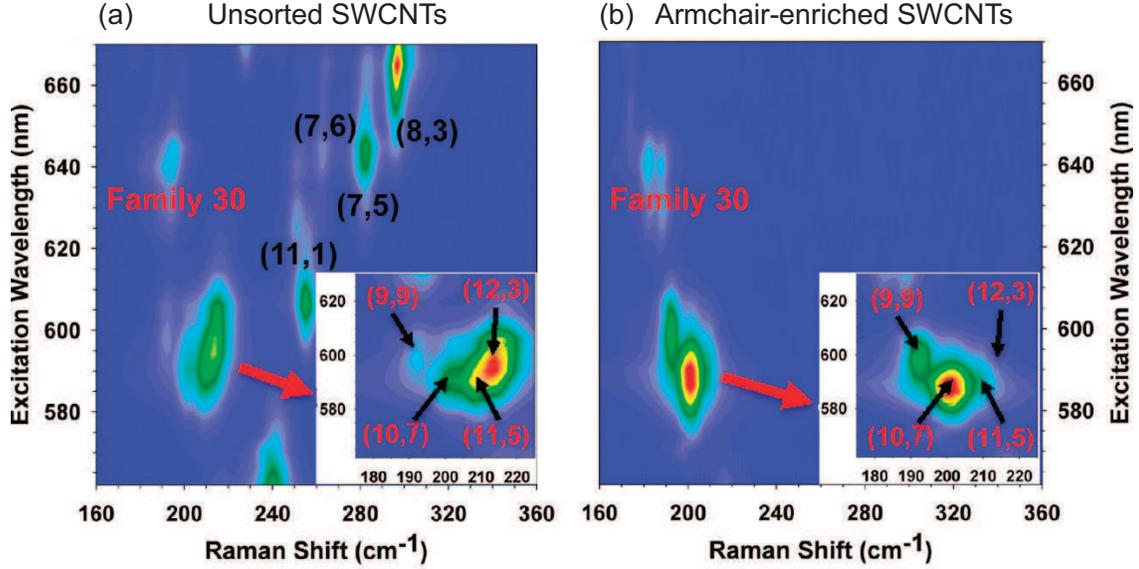}
\caption{Resonant Raman scattering RBM intensity contour plots of (a) unsorted and (b) armchair-enriched HiPco SWCNT samples taken over an excitation range of 562-670 nm. The insets of (a) and (b) highlight family 2$n+m$~=~27. The scale of the inset of (a) has been magnified by 1.5 relative to the full excitation map to differentiate the ($n,m$) members of family 27 more clearly.   Raman intensity excitation maps of (b) clearly show enrichment of armchair and near-armchair species relative to (a).  Reproduced from Reference \onlinecite{HarozetAl10ACS}.}
\label{fig-RamanmapsG}
\end{figure*}

Figure \ref{fig-RamanmapsG} shows two-dimensional contour plots of Raman intensity corresponding to the RBM region as a function of Raman shift and excitation wavelength (562-670 nm) of unsorted (Fig.~\ref{fig-RamanmapsG}a) and metal-enriched (Fig.~\ref{fig-RamanmapsG}b) HiPco material.  Each unique intensity feature corresponds to the RBM of a particular ($n,m$) species.  In this particular excitation region for HiPco material, RBM features correspond to members of the metallic 2$n$+$m$ families 27 and 30 as well as small-diameter semiconductors of families 19, 20 and 23.  In comparing both Raman ``maps", immediately it is observed that the RBMs corresponding to semiconductors are completely missing, confirming observations by optical absorption (Section \ref{absorption}).  However, more significant and surprising is the change in relative RBM intensity of species of metallic family 27.  As can be most clearly observed in the insets of Fig.~\ref{fig-RamanmapsG}, RBM intensity is highest for the (12,3) and (11,5) species in the unsorted sample, but the (9,9) and (10,7) species are most intense in the metal-enriched sample, with the (11,5) and (12,3) species virtually absent.    This change in relative intensity of the same family of RBMs between the unsorted and metal-enriched samples suggests that DGU may be affecting the chiral distribution of species, favoring large chiral angle, $\nu$~=~0 species.  

\begin{figure*}
\centering
\includegraphics[scale=.6]{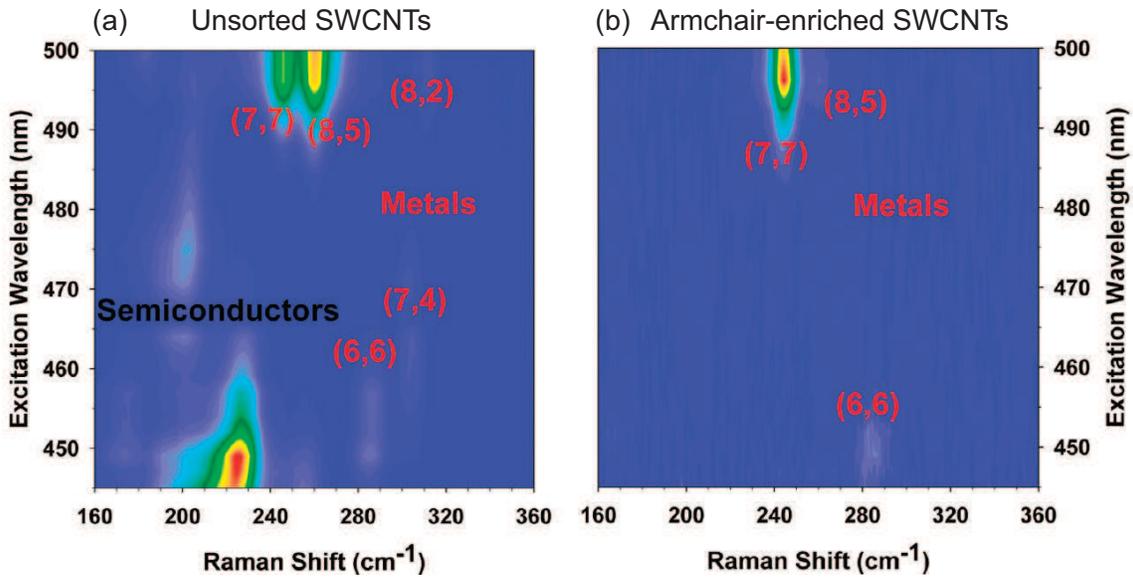}
\caption{Resonant Raman scattering RBM intensity contour plots of (a) unsorted and (b) metal-enriched HiPco SWCNT samples taken over an excitation range of 440-500 nm. Again, Raman intensity excitation maps of (b) clearly show enrichment of armchair carbon nanotubes relative to (a), this time for families 21 and 18.  Reproduced from Reference \onlinecite{HarozetAl10ACS}.}
\label{fig-RamanmapsB}
\end{figure*}

To explore this behavior further, other 2$n$+$m$ families can be examined by tuning to shorter excitation wavelengths.  Figure \ref{fig-RamanmapsB} displays Raman spectra taken continuously from 440-500 nm for unsorted and metal-enriched samples.  In the unsorted sample (Fig.~\ref{fig-RamanmapsB}a), we observe RBMs from multiple members of $\nu$~=~0 families 21 and 18, excited via their $E_{11}^{M-}$ optical transition, and at lower frequencies, semiconductors, excited via their $E_{33}^{S}$ optical transition.  However, in the metal-enriched sample (Fig.~\ref{fig-RamanmapsB}b), not only have the RBMs corresponding to semiconducting species vanished but even more stunningly, the non-armchair members of families 21 and 18 (i.e.,~(8,5) and (7,4) \& (8,2), respectively) have also diminished substantially, leaving behind only the armchair species (7,7) and (6,6).  Similar behavior is observed for Raman data probing metallic family 24 (not shown), where species (11,2), (10,4), (9,6) and (8,8) are present in the unsorted sample but are all absent except (8,8) and a trace amount of (9,6) in the metal-enriched data.  

\begin{figure*}
\centering
\includegraphics[scale=.9]{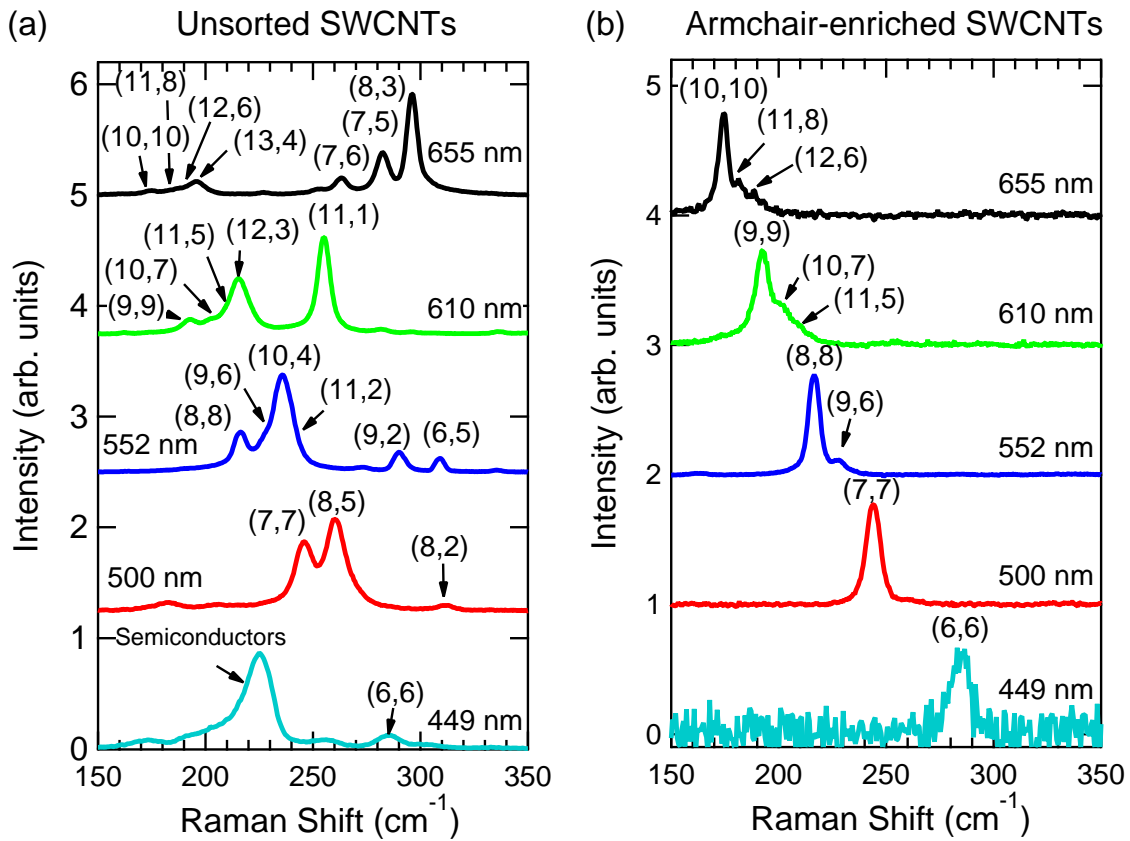}
\caption{Selected Raman spectra taken from data contained within Fig.~\ref{fig-RamanmapsG} and \ref{fig-RamanmapsB} comparing (a) before and (b) after DGU, showing evidence for the enrichment of (a) ``metallic'' $\nu$~=~0 nanotubes, and in particular, (b) armchair, or ($n,n$), carbon nanotubes. Adapted from Reference \onlinecite{HarozetAl10ACS}.}
\label{fig-ArmchairRBM2}
\end{figure*}

To summarize Figs.~\ref{fig-RamanmapsG} and~\ref{fig-RamanmapsB}, we can examine Raman spectra taken at selected excitation wavelengths for both unsorted and metal-enriched HiPco samples.  Figure \ref{fig-ArmchairRBM2} displays RBM Raman spectra taken at 655, 610, 552, 500, and 449 nm excitation wavelength for unsorted (Fig.~\ref{fig-ArmchairRBM2}a) and metal-enriched (Fig.~\ref{fig-ArmchairRBM2}b) samples.  Strikingly, all RBMs corresponding to semiconductors are completely suppressed as well as many of the low-to-medium chiral angle $\nu$~=~0 species.  Only $\nu$~=~0 species of large chiral angle remain, with the strongest enrichment towards the ($n,n$) or armchair species.  The fact that all armchair species are enriched after DGU, while only some of the large chiral angle, $\nu$~=~0 species persist, suggests that the DGU mechanism is dependent not only on electronic type but also on molecular structure dependence specifically involving chiral angle.  It should be noted as well that all measured RBMs for non-armchair, $\nu$~=~0 SWCNTs are due to resonances with $E_{11}^{M-}$ transitions.  Resonance with the $E_{11}^{M+}$ transitions was not observed, consistent with results by Son \textit{et al.}\cite{SonetAl06PRB}~and Doorn \textit{et al.}\cite{Doornetal08PRB} for small-diameter SWCNTs.

While the change in relative Raman intensity of RBMs between unsorted and sorted samples is a clear indicator that armchair enrichment is occurring, accurate quantitative statements of enrichment can be problematic to make based on the difficulties inherent to RRS in SWCNTs.  Explicitly, SWCNT RRS intensity, $I_{RRS}(n,m)$, as described by
\begin{equation}
I_{\rm RRS}(n,m) \propto N \left| \frac{M_{op}^{~2}~M_{e-ph}}   {(E_{L} - E_{ii} - \Gamma _{el})(E_{L} - E_{ii} - \hbar \omega _{ph} - \Gamma _{el})} \right|^{2}
\label{RRS eqn}
\end{equation}
is a function of the excitation laser energy, $E_{L}$, the optical transition energy, $E_{ii}$, the phonon energy, $\hbar \omega_{ph}$, the electronic broadening factor or optical transition line width, $\Gamma_{el}$, the exciton-photon coupling matrix element, $M_{op}$, the exciton-phonon coupling matrix element, $M_{e-ph}$, and the abundance of that particular ($n,m)$ species, $N$.  The use of single-excitation-wavelength Raman data to determine the degree of enrichment of particular species complicated due to the fact that different species resonate maximally at different excitation wavelengths $E_{ii}$ (i.e., there is no unique excitation wavelength that resonates equally well with all species).  Although relative abundances cannot be determined from such data, an argument can be made for the use of single-excitation-wavelength Raman data as a preliminary screening to indicate that a change in relative abundance may be occurring.  This assumes that the resonant condition for a particular ($n,m$) species has not changed (optical transition energy and line width) due to the separation process (e.g., doping, aggregation, functionalization, etc.).  For a more fair comparison between species within the same sample and between different samples, Raman data collected continuously over a broad excitation energy range ensures that each ($n,m$) species is sampled at its peak intensity producing a numerical intensity proportional to abundance.\cite{JorioetAl05PRB2}  However, this too can still be problematic as the peak ``intensity'' must be carefully interpreted, especially to account for the chiral-dependent optical transition line width, $\Gamma_{el}$, as well as the RBM phonon line width, $\Gamma_{ph}$, which can affect the observed peak height.  The use of RBM peak area as opposed to RBM peak height as the quantity of comparison avoids the latter issue.  

\begin{figure}
\centering
\includegraphics[scale=.7]{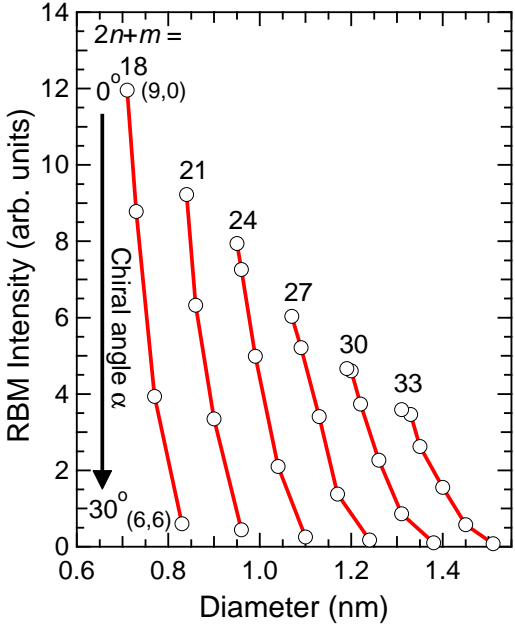}
\caption{Calculated RBM Raman intensities\cite{SatoetAl10CPL} for $\nu$~=~0 SWCNT species as a function of nanotube diameter and chiral angle, with members of the same 2$n$+$m$ family grouped together.  Clearly, RBM Raman intensity increases with decreasing nanotube diameter and chiral angle with armchair (zigzag) species having the smallest (largest) RBM intensities for a given 2$n$+$m$ family.}
\label{fig-TheoRBMInt}
\end{figure}

A more representative analysis uses the RBM ``intensity" for a given ($n,m$) species as a function of excitation energy, the so-called Raman excitation or resonance profile (REP).  By measuring a complete REP for a given species, the integrated RBM area for each ($n,m$) across the entire range of excitation can be calculated, which accounts for all the aforementioned issues.\cite{JorioetAl05PRB2}  Finally, to determine relative chiral abundances, the ($n,m$)-dependent Raman cross-sections must be considered and corrected.  Jiang \textit{et al.}\cite{JiangetAl07PRB2} and later Sato \textit{et al.}\cite{SatoetAl10CPL} calculated the exciton-photon and exciton-phonon coupling matrix elements for many ($n,m$), considering contributions of curvature, excitons and other many-body effects, to explain the chiral dependence of Raman intensity.  While $M_{op}$ possesses a very weak dependence on chiral angle and increases in magnitude with decreasing nanotube diameter, $M_{e-ph}$ has a very strong dependence on chiral angle with increasing coupling occurring with decreasing chiral angle.  In fact, the zigzag or ($3n,$0) species can possess as much as an order of magnitude larger $M_{e-ph}$ than the armchair species, for the same $2n$+$m$ family.\cite{JiangetAl07PRB2, MachonetAl05PRB, GoupalovetAl06PRB}  Combining both matrix element behaviors together, Jiang \textit{et al.}~and Sato \textit{et al.}~observe (as shown in Fig.~\ref{fig-TheoRBMInt}) an increase in Raman intensity $I_{RRS}$ with decreasing diameter and decreasing chiral angle for metallic species, illustrating the need for correcting experimentally determined, integrated Raman intensities for ($n,m$)-dependent Raman cross-sections to compare each chiral species fairly for the purposes of measuring relative abundances.

\begin{figure}
\centering
\includegraphics[scale=.85]{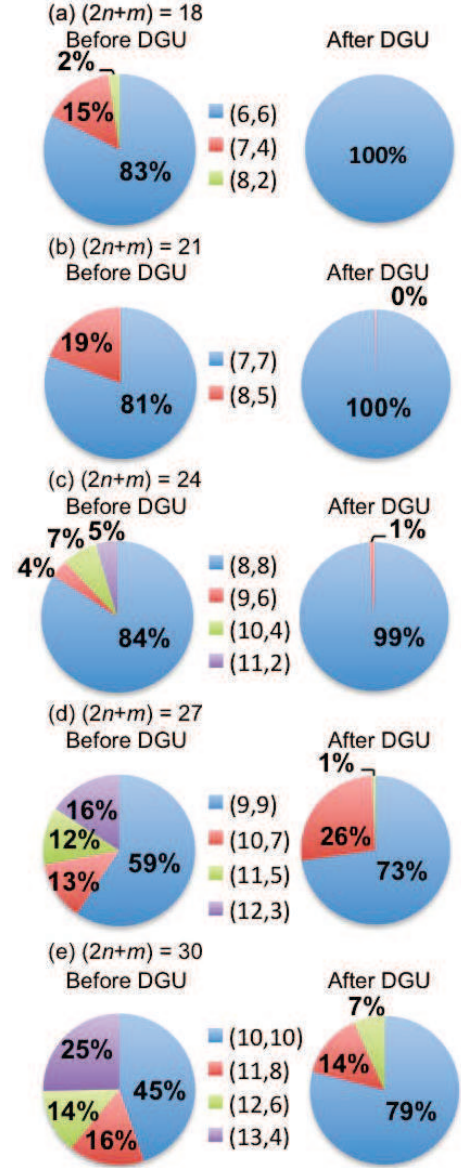}
\caption{Comparison of the relative abundances of $\nu$~=~0 species in HiPco SWCNT material, before DGU (unsorted) and after DGU (armchair-enriched), as determined by resonant Raman scattering.  ($n,m$) species are grouped together by 2$n$+$m$ family, (a)-(e) corresponding to families 18, 21, 24, 27 and 30, respectively.}
\label{fig-piechart5}
\end{figure}

Taking into account all the aforementioned considerations, we used the data contained in Figs.~\ref{fig-RamanmapsG} and \ref{fig-RamanmapsB} and Reference \onlinecite{HarozetAl10ACS} to construct REPs of the RBM for each metallic species in the unsorted and metal-enriched HiPco samples, which were then subsequently used to calculate relative abundance.  Specifically, we fitted Raman spectra taken at each excitation wavelength as a sum of Lorentzian peaks, with each peak corresponding to an observed RBM.  This first iteration of fits allowed peak position, line width, and area to vary to allow the most unbiased fitting result.  After this first iteration of fitting, the fitting parameters for each RBM, taken at its resonant Raman maximum wavelength, were used to refit all other spectra for the same RBM to further refine fitting allowing only peak area to vary.  From these refined fits, a REP was constructed for each ($n,m$) as a function of excitation energy using RBM peak area.  The resulting REPs were then fitted to a modified version of Eq.~(\ref{RRS eqn}), where the numerator terms were combined into a single quantity $A$, amplitude, which includes contributions from $M_{op}$, $M_{e-ph}$, and $N$.  These REP fits were then integrated over excitation energy to produce an integrated Raman ``intensity" (area) for each RBM.  This integrated REP area represents the total Raman signal contribution for a particular ($n,m$) over the entire excitation region and should be proportional to its particular abundance.  Each integrated ($n,m$) area can then be corrected by dividing it by the corresponding theoretical Raman cross-section value calculated by Sato \textit{et al.}\cite{SatoetAl10CPL}, finally providing us with the relative abundance of each chiral species.  Figure~\ref{fig-piechart5} summarizes the results of this entire analysis in the form of pie charts representing the percent abundance of each member of a given $2n$+$m$ before and after DGU separation.   
%
%
%

Immediately apparent in the comparisons of the before and after DGU populations in Fig.~\ref{fig-piechart5} is that the metal-enriched (after DGU) sample is in fact strongly enriched towards armchair species as compared to the unsorted (before DGU) sample with small chiral angle species having been eliminated from the enriched sample (Note: In Fig.~\ref{fig-piechart5}b, the (9,3) species was detected only at a single excitation wavelength (514.5~nm) before DGU, however, due to insufficient data, an REP could not be constructed and as such relative abundance could not be determined).  A closer examination of the species remaining in the ``armchair-enriched" sample indicates that any metallic species with a chiral angle smaller than $\approx$23-24$^{\circ}$ is significantly suppressed by the DGU process, indicating that the three-surfactant DGU method has a critical chiral angle below which enrichment does not occur and suggesting the possible role of chiral-angle-specific steric interactions between chiral surfactants, such as DOC and SC, and a particular ($n,m$) nanotube geometry.\cite{GreenetAl09NanoRes, HarozetAl10ACS}  

Further evidence for this effect is suggested by the relative abundances of species within the unsorted sample which is in aqueous suspension by DOC.  Raman measurements by Fantini \textit{et al.}\cite{FantinietAl04PRL} and Maultzsch \textit{et al.}\cite{MaultzschetAl05PRB} on aqueous dispersions of HiPco material suspended by the achiral surfactant SDS indicate that all members of a particular metallic $2n$+$m$ family, including zigzag species, are present.  However, examination of the chiral angles of species present in our unsorted, DOC-suspended sample points out that the zigzag and near-zigzag species are absent after the initial suspension process.  The smallest chiral angle species present is the (11,2) and the largest chiral angle species not present is that of the (14,2), suggesting that species with a chiral angle less than 6-8$^{\circ}$ are not suspended by DOC.  Such selective geometrical interactions in SWCNTs are not new in the literature (for example, Ju \textit{et al.}\cite{JuetAl08NatureNano}) but previously unreported in this commonly used surfactant and indicates that caution should be taken with statements regarding the composition of a particular nanotube sample as being ``as-produced".  

Another interesting observation is the on-average larger abundance of large chiral angle species already present in the unsorted sample.  In particular, armchair species seem to be more abundant than other species, in agreement with recent theoretical results by Ding \textit{et al.}\cite{DingetAl09PNAS} and experimental results by Rao \textit{et al.}\cite{RaoetAl12NatureMat} that suggest that armchair and other large chiral angle nanotube species kinetically grow faster than zigzag and other small chiral species.  Although, the initial goal of this population study was merely to indicate whether the DGU process affected the chiral distribution of species within a given SWCNT material, the above data and analysis point to other fruitful avenues of investigation with regards to type- and chiral-separation mechanisms, nanotube-surfactant interactions, and nanotube growth kinetics.  



\section{G-band Raman Signatures of Armchair and Non-armchair $\nu = 0$ Nanotubes}
\label{RRS-Gband}

Utilizing the high-purity, armchair-enriched samples described in Sections \ref{Separation}-\ref{RRS-RBM}, other important Raman modes relevant to optical relaxation and electronic scattering processes can be studied to examine intrinsic line shape and its relationship to nanotube chiral structure.      
Of particular interest is the G-band, a Raman-active optical phonon feature due to the in-plane C-C stretching mode of $sp^{2}$-hybridized carbon.  Although a single frequency feature in monolayer graphene,\cite{FerrarietAl06PRL} the graphitic parent material of $sp^{2}$-hybridized carbon, in SWCNTs, the G-band splits into two, the higher-frequency G$^{+}$ and lower-frequency G$^{-}$ peaks, due to the curvature-induced inequality of the two bond-displacement directions.  For $\nu$~=~0 tubes, the G$^{+}$ mode is a narrow Lorentzian peak, while the G$^{-}$ mode is extremely broad.  Earlier theoretical studies described this broad feature as a Breit-Wigner-Fano line shape due to the coupling of phonons with an electronic continuum\cite{BrownetAl01PRB} or low-frequency plasmons,\cite{Kempa02PRB} but there is now growing consensus that the broad G$^{-}$ peak is a frequency-softened and broadened longitudinal optical (LO) phonon feature, a hallmark of Kohn anomalies.\cite{DubayetAl02PRL, LazzerietAl06PRB, IshikawaAndo06JPSJ, PiscanecetAl07PRB, NguyenetAl07PRL, WuetAl07PRL, FarhatetAl07PRL}  Through either scenario, this broad G$^{-}$ peak has conventionally been known to be a ``metallic'' feature, indicating the presence of metallic tubes.

Single-nanotube Raman measurements of the G-band taken on armchair SWCNTs have produced a variety of observations with both the absence\cite{WuetAl07PRL,MicheletAl09PRB,BerciaudetAl10PRB} and presence\cite{ParketAl09PRB} of a broad G$^{-}$ feature recorded for freestanding and substrate-supported individual SWCNT samples, respectively.  Data taken on freestanding, single-tubes in references,\cite{WuetAl07PRL,MicheletAl09PRB,BerciaudetAl10PRB} where ($n$,$m$) is assigned independently by other means, as compared to the more commonly used substrate-supported samples is mainly the source of the variability in the interpretation of the G$^{-}$ peak.    This inconsistency most likely stems from the sensitivity of G-band phonons to doping and charge transfer from intentional gating\cite{NguyenetAl07PRL,WuetAl07PRL} and environmental effects (common to substrate-supported single SWCNTs) that can induce a localization of electron density in the nanotube $\pi$~electron cloud\cite{RaoetAl97Nature,ShimetAl08JPCC} and which ultimately modify SWCNT band structure.  An excellent example of this sensitivity to environment is demonstrated by Shim \textit{et al.}\cite{ShimetAl08JPCC} where Raman spectra of multiple examples of substrate-supported (12,6) resulted in widely differing G-band structure with regards to G$^{-}$ intensity, line shape  and line width.  As a consequence, great care must be taken in generalizing observed G-band structure at the single-tube level for a given chiral species because of this sensitivity and the fact that only a few nanotube examples are typically sampled in an experiment.  

To avoid the experimental difficulties and uncertainties of previous single-nanotube Raman measurements, macroscopic ensembles enriched in armchair and bulk $\nu$~=~0 ``metallic" species can be used to more clearly elucidate the intrinsic G-band line shape as a function of nanotube structure.  Using detailed wavelength-dependent Raman scattering measurements collected on macroscopic ensembles\cite{HarozetAl11PRB, TuetAl11JACS} enriched via density gradient ultracentrifugation (armchair-enriched, AE-SWCNT)\cite{HarozetAl10ACS} and (``metal"-enriched, ME-SWCNT)\cite{NiyogietAl09JACS} and DNA-ion exchange chromatography (single-chirality armchair, DNA-SWCNT)\cite{TuetAl11JACS} methods, spectra measure the average Raman response from $\approx$10$^{10}$ tubes, removing the issue of inconsistencies caused by tube-to-tube variations due to structural defects and inhomogeneous environment.  As a reference, all samples were compared to unsorted SWCNTs (AP-SWCNT) suspended aqueously in DOC by the standard ultracentrifugation method.\cite{OConnelletAl02Science}  

\begin{figure}
\centering
\includegraphics[scale=.53]{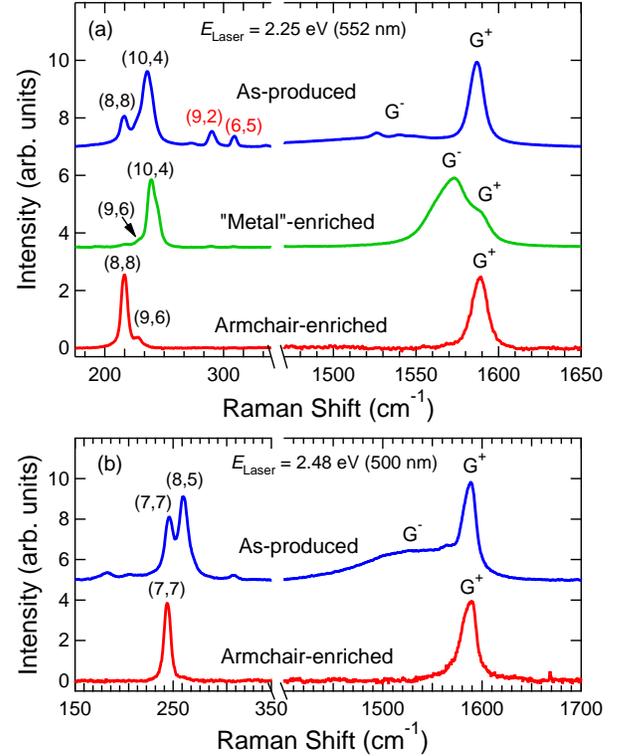}
\caption{Raman spectra of RBM and G-band for AP-SWCNT, ME-SWCNT, and AE-SWCNT samples taken at excitation wavelengths (a) 552~nm and (b) 500~nm where SWCNTs from families 2$n$+$m$~=~24 and 21, respectively, are primarily probed.  In both sub-figures, appearance of the broad G$^{-}$ feature corresponds to resonance with non-armchair species such as (10,4) in ME-SWCNT in (a) and (8,5) in AP-SWCNT in (b).  In the case of singular resonance with armchair species [(8,8) and (7,7)], a single and narrow G$^{+}$ peak is observed.  Reproduced from Reference \onlinecite{HarozetAl11PRB}.}
\label{fig-Gbandmixed}
\end{figure}

Figure \ref{fig-Gbandmixed}a shows resonant Raman spectra for both the RBM and G-band ranges for AP-SWCNT, ME-SWCNT, and AE-SWCNT samples, collected at 552~nm laser excitation.  Based on previous Raman measurements and predictions based on Kataura plots, this wavelength primarily resonates with members of the 2$n$+$m$~=~24 family [(8,8) and (10,4)] as well as some small-diameter $\nu$~=~1 species [(9,2) and (6,5)].  Examining the RBM region first, ME-SWCNT (green trace) and AE-SWCNT (red trace) samples are clearly enriched in $\nu$~=~0 tubes while the AP-SWCNT sample (blue trace) contains a mixture of both $\nu$~=~1 and $\nu$~=~0 species.  Furthermore, the ME-SWCNT sample contains most of the members of family 24, whereas the AE-SWCNT sample contains only the (8,8) armchair species and a very small amount of the non-armchair species (9,6).

In the corresponding G-band region for the AP-SWCNT (blue trace) in Fig.~\ref{fig-Gbandmixed}a, we observe a G-band line shape typical of resonance with $\nu$~=~1 species such as (9,2) and (6,5),
because the intensity of the G$^{+}$ peak for $\nu$~=~$\pm$1 tubes is markedly stronger than that for $\nu$~=~0 tubes.\cite{MachonetAl06PRB,JiangetAl07PRB}  With the removal of the $\nu$~=~$\pm$1 impurities, the ME-SWCNT sample displays a G-band line shape consistent with what is commonly attributed in the literature to the presence of only metallic nanotubes with the notably broad and intense G$^{-}$ component (green trace in Fig.~\ref{fig-Gbandmixed}a), centered at $\sim$1573~cm$^{-1}$.  However, the AE-SWCNT sample, which is also free of $\nu$~=~$\pm$1 impurities, displays a completely different G-band line shape consisting of only a single and narrow G$^{+}$ component (red trace in Fig.~\ref{fig-Gbandmixed}a), centered at $\sim$1589~cm$^{-1}$ and consistent with the G$^{+}$ component of the ME-SWCNT sample.  Such a large difference in G-band line shape between these two samples suggests a significant ($n$,$m$)-dependence of the G$^{-}$ feature.  In fact, when taken together with the RBM spectrum for the ME-SWCNT sample, a correlation between the appearance of small chiral-angle (or zigzag-like) $\nu$~=~0 species and that of the G$^{-}$ peak is apparent.  Furthermore, this peak becomes absent at resonance with armchair tubes.

Examining another $\nu$~=~0 family, 2$n$+$m$~=~21, we see a similar trend.  In Fig.~\ref{fig-Gbandmixed}b, resonant Raman spectra of the RBM and G-band regions for the AP-SWCNT and AE-SWCNT samples taken at 500~nm excitation where both samples are spectrally isolated from any $\nu$~=~$\pm$1 impurities, and, as such, only resonate with $\nu$~=~0 tubes.  The AP-SWCNT sample displays resonance with only two species, namely (8,5) and (7,7) corresponding to the  observation of the broad G$^{-}$ feature and a narrow G$^{+}$ component.  In contrast, the AE-SWCNT sample resonates singularly with the armchair species (7,7) and again displays only a single, narrow G$^{+}$ peak.

\begin{figure}
\centering
\includegraphics[scale=.45]{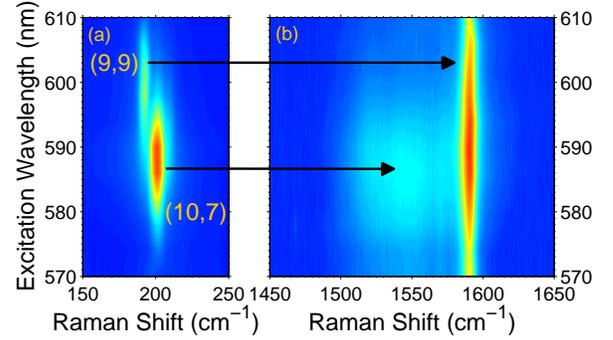}
\caption{Raman intensity for AE-SWCNT taken as a function of Raman shift and excitation wavelength for family 2$n$+$m$~=~27.  (a) RBM region where two clear RBMs due to the (9,9) and (10,7) are observed.  (b) Corresponding G-band region where only the G$^{+}$ peak is observed when resonating primarily with the (9,9) and the appearance of the broad G$^{-}$ coincides with the maximum of the (10,7) RBM.  Reproduced from Reference \onlinecite{HarozetAl11PRB}.}
\label{fig-GbandContour}
\end{figure}

To further examine the relationship between the appearance of the broad G$^{-}$ peak and ($n$,$m$) composition, Raman intensity of another isolated $\nu$~=~0 family, 2$n$+$m$~=~27, as a function of both Raman shift and excitation wavelength was examined.  Figure \ref{fig-GbandContour} shows a contour plot of Raman intensity of the AE-SWCNT sample for the RBM and G-band regions over excitation wavelengths of 570-610~nm.  As is seen in Fig.~\ref{fig-GbandContour}(a), only two RBMs, (9,9) and (10,7), are observed again reflecting the enrichment toward armchair-like species of the AE-SWCNT sample.  The corresponding G-band [Fig.~\ref{fig-GbandContour}(b)] shows only a single G$^+$ peak centered at $\sim$1590~cm$^{-1}$ at the longest excitation wavelengths ($\sim$610~nm) coinciding with the resonance maximum of the (9,9) RBM.  As the excitation wavelength is decreased, a second, broad Raman feature appears at lower frequency alongside the 1590~cm$^{-1}$ peak.  Centered at $\sim$1550~cm$^{-1}$, this feature reaches a maximum in intensity at $\sim$587~nm, which coincides with the RBM resonance maximum of the (10,7).  Therefore, we assign the 1550 and 1590~cm$^{-1}$ features to the G$^{-}$ and G$^{+}$ peaks of the G-band.  The simultaneous appearance of the G$^{-}$ peak with the Raman resonance maximum of the (10,7) and its absence with pure resonance with the (9,9) point out a clear correlation between ($n$,$m$) chirality and G-band line shape.  Namely, the broad G$^{-}$ peak appears only in the presence of non-armchair $\nu$~=~0 nanotube species, and ($n$,$n$) armchair species consist of only one single, narrow peak.

\begin{figure}
\centering
\includegraphics[scale=.55]{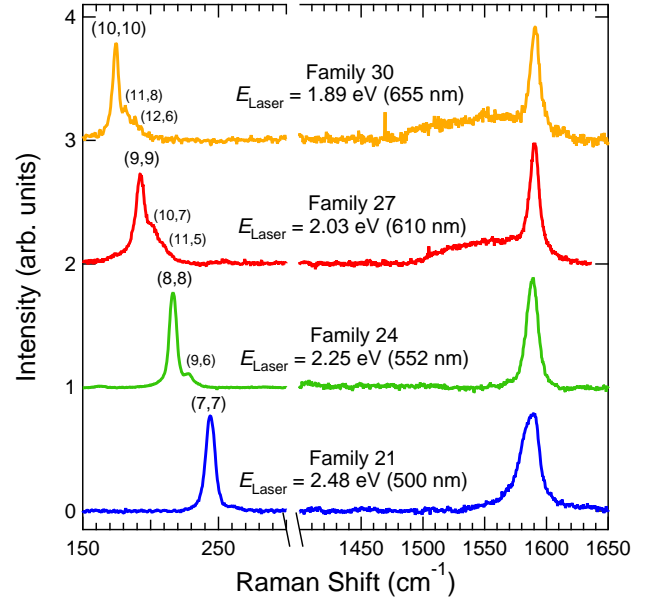}
\caption{Selected resonant Raman spectra for AE-SWCNT taken at excitation wavelengths 655-, 610-, 552-, and 500-nm, where resonance primarily occurs with armchair species (10,10), (9,9), (8,8), and (7,7), respectively.  In each case, the G-band reflects contribution mainly from the G$^{+}$ peak only.  Reproduced from Reference \onlinecite{HarozetAl11PRB}.}
\label{fig-GbandArmchairErik}
\end{figure}


\begin{figure}
\centering
\includegraphics[scale=.55]{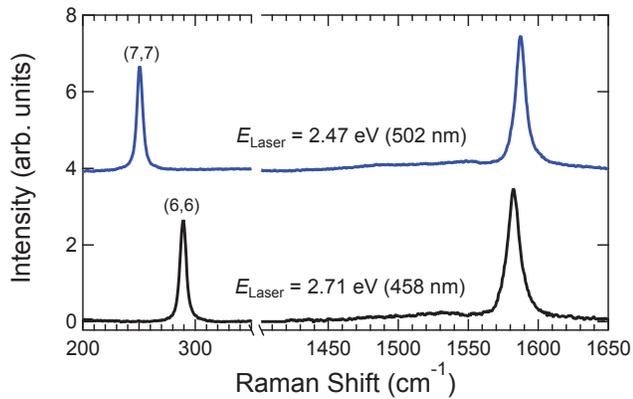}
\caption{Resonant Raman spectra for single-chirality armchair SWCNT samples produced by the DNA-based IEX method, taken at excitation wavelengths 502- and 458-nm.  Because of the extremely singular purity of these samples, a single G$^{+}$ peak is observed due to the armchair species (7,7) and (6,6), respectively. Adapted from Reference \onlinecite{TuetAl11JACS}.}
\label{fig-GbandArmchairMing}
\end{figure}

Finally, to examine the generality of this behavior to other nanotube families, Fig.~\ref{fig-GbandArmchairErik} shows selected Raman spectra of the (7,7), (8,8), (9,9), and (10,10) armchair families taken at 500, 552, 610, and 655~nm excitations, respectively, for the AE-SWCNT sample.  Figure~\ref{fig-GbandArmchairMing} displays Raman spectra examining separate pure, single-chirality (6,6) and (7,7) samples (DNA-SWCNTs) taken at 502 and 458~nm excitations, respectively.  These wavelengths are close to the general resonance maxima for each respective family.  Because these data span a broad diameter range of 0.83-1.38~nm and more importantly across 5 different armchair families, the appearance and dominance of a single G$^{+}$ feature in Raman spectra of $\nu$~=~0 species is a general result and indicator for the predominance of armchair species in a SWCNT sample.  The observation of the weak G$^{-}$ peak at 610 and 655~nm excitation in Fig.~\ref{fig-GbandArmchairErik} represents the non-negligible but small contribution of residual non-armchair $\nu$~=~0 impurities [(10,7) and (11,8)] and a qualitative measure of the larger Raman cross-section such species have as compared to armchair tubes.\cite{JiangetAl07PRB}  

Another interesting observation to make is that of the larger line width of the G$^{+}$ feature observed for the (7,7) spectrum taken at  500~nm in Fig.~\ref{fig-GbandArmchairErik} than that of the (7,7) sample taken at 502~nm in Fig.~\ref{fig-GbandArmchairMing}.  Although environmental factors can play a role in affecting peak position and line width to some degree (the AE-SWCNT sample is ultimately suspended in DOC (H$_{2}$O) and the DNA-SWCNT samples are suspended by DNA (H$_{2}$O)), the larger line width of the AE-SWCNT (7,7) spectrum ($\omega_{G}\approx$~1587~cm$^{-1}$,~$\Delta\omega\approx$~76~cm$^{-1}$) as compared to the DNA-SWCNT (7,7) spectrum ($\omega_{G}\approx$~1587~cm$^{-1}$,~$\Delta\omega\approx$~21~cm$^{-1}$) suggests that additional frequency contributions are responsible for the broadening observed in the G$^{+}$ feature of AE-SWCNT (7,7).  This may be due to a combination of G-band contributions from neighboring $2n$+$m$ semiconducting families (which are absent in the pure (7,7) DNA-SWCNT sample) for which the G-band contribution can still be noticeable despite undetectable RBMs and the very large resonance window of the G-band when considering both incident and scattered Raman resonances ($\geq$~250 meV).  Further investigation is warranted to determine the origin of this broadening but does illustrate the power of the G-band for assessing chiral purity of enriched samples based on the large resonance window and sensitivity of the G-band to detect species not observed via the RBM.

Based on the aforementioned results, we attribute the narrow G$^{+}$ component observed during resonance with all $\nu$~=~0, armchair and non-armchair species to the transverse optical (TO) phonon mode and the broad G$^{-}$ component observed only with $\nu$~=~0, non-armchair species to the LO phonon mode arising from phonon softening due to Kohn anomaly.\cite{DubayetAl02PRL,LazzerietAl06PRB,IshikawaAndo06JPSJ,PiscanecetAl07PRB,NguyenetAl07PRL,WuetAl07PRL,FarhatetAl07PRL}  The absence of the LO phonon feature for armchair tubes is consistent with some theoretical studies (e.g., References \onlinecite{DubayetAl02PRL} and \onlinecite{SaitoetAl01PRB}), i.e., it is {\em not} Raman active in armchair SWCNTs (although this still does not eliminate the presence of a Kohn anomaly- only its Raman activity), while it is inconsistent with more recent theories.\cite{SasakietAl08PRB,ParketAl09PRB}

In conclusion, using resonant Raman measurements of macroscopic ensembles of $\nu$~=~0 enriched SWCNT samples, we have demonstrated that the appearance of a broad G$^{-}$ peak is due to the presence of non-armchair ``metallic'' tubes.  More importantly, the G-band of the truly gapless armchair tubes consists of a single, narrow G$^{+}$ component and serves as a further diagnostic for their identification.  Because of the large statistical sampling presented here, these results are generalizable to all $\nu$~=~0 nanotubes. Finally, this study demonstrates the ability of using specialized, enriched samples to extract meaningful and definitive information about single chiralities from a macroscopic scale, enabling future studies beyond the experimental limitations of the single-tube level.

\section{Summary and Outlook}
\label{outlook}
We have highlighted a few of the fascinating properties in metallic and narrow-gap single-wall carbon nanotubes using optical spectroscopies.  Using post-growth separation techniques (DGU and DNA-based methods), we have shown that it is possible to create macroscopic ensembles of armchair carbon nanotubes.  In the case of DGU-based separation, we were able to enrich SWCNT material in armchair carbon nanotubes over a broad range of diameters and different starting materials.  By measuring the Raman excitation profiles of the unique radial breathing mode for each ($n,m$), we determined the relative populations of each species before and after enrichment proving that DGU enriches largely toward armchair carbon nanotubes.  Using that population information, we were able to show through fits to optical absorption spectra that although armchair SWCNTs are metallic in conduction behavior, they display highly symmetric absorption line shapes indicative of the considerable contribution excitons play in the optical processes of these one-dimensional metals.  Using DNA-based separation, single-chirality armchair ensembles of (6,6) and (7,7) were prepared that revealed more fine structure including a clearly observable, high-energy peak attributed to a phonon sideband, in addition to the previously observed discrete exciton and continuum features.  Using both preparations of enriched samples, clear measurements of the G-band line shape via resonant Raman scattering showed that the broad G$^{-}$ feature, once attributed to resonance with metallic species, is in fact only observed with non-armchair, $\nu$~=~0 nanotubes and is completely absent with resonance with solely armchair SWCNTs.  

Now armed with these types of highly armchair-enriched, macroscopic samples, other optical phenomena and properties can be probed experimentally.  For instance, while in-depth studies have been conducted of the excitation dependence of the RBM frequency in SWCNTs, in metallic nanotubes, it is usually impossible in ensemble samples to distinguish the individual component resonances due to the incident and scattered photons.  This is due to the small RBM phonon energy separating resonances and the large line-width of each resonance (i.e., $\hbar \omega_{\rm RBM} < \Gamma_{\rm el}$).\cite{SatishkumaretAl06PRB}  However, using the G-band phonon of an armchair nanotube, where $\hbar \omega_{\rm G^{+}} > \Gamma_{\rm el}$, should allow clear observation of each resonance.  In particular, asymmetry in the excitation profile of the G-band due to non-Condon effects is of interest, as has been recently observed for single-chirality semiconductor samples.\cite{DuqueetAl11ACS}  Another phonon mode of interest is the G$'$-band (also referred to as the 2D band).  Because of its highly dispersive nature through a double resonance process, measurement of its excitation profile should reveal significant information about the dispersion of the optical transition through which it is resonant.  In optical absorption, single-chirality armchair samples can provide measurement of the absorption cross-section of a given armchair species, a fundamental quantity necessary for quantitative estimates of exciton-photon coupling matrix elements as well as nanotube concentration.  More exciting is the potential to study higher and lower energy features such as the M-point singularity (sometimes referred to in the literature as the $\pi$-plasmon) in the ultraviolet and the broad 4 THz feature observed by many groups in the far-infrared,\cite{UgawaetAl99PRB, ItkisetAl02NL, JeonetAl02APL, AkimaetAl06AM, BorondicsetAl06PRB, NishimuraetAl07APL, KampfrathetAl08PRL} respectively, via polarized absorption.  Still more exotic is the opportunity to study degenerate one-dimensional electrons through temperature- and magnetic-field-dependent optical- and terahertz-frequency absorption with aligned armchair SWCNT thin films, searching for many-body phenomena such as Fermi-edge singularity and Tomonaga-Luttinger liquid behavior.  Finally, probing further to even lower frequencies (GHz), studies on pure, single-chirality armchair samples via microwave electron spin resonance may allow observation of spin-charge separation, a hallmark of Luttinger liquid behavior.  Future studies on single-chirality narrow-gap semiconductor SWCNT samples should provide enormous insight into the role upper-branch ``metallic" transitions play in absorption processes and whether interference effects between upper and lower branches occur.  From an applications perspective, aligned, macroscopic films of $\nu$~=~0 species have enormous potential to serve as THz-frequency detectors and polarizers\cite{RenetAl09NL, RenetAl12NL} due to their highly anisotropic optical properties combined with their excellent electrical conductivity,\cite{TansetAl97Nature, WhiteTodorov98Nature} bridging the technology gap between optics and electronics.  Lastly, although yields from separation processes like DGU and DNA-based separation are low, such separated materials could serve as seed crystals for nanotube cloning where existing nanotubes are forced to restart growth to produce more material of the parent armchair chirality, allowing researchers to eventually produce gram quantities of armchair SWCNTs.  

Taken together, these results and ongoing research efforts paint a bright and fruitful future for carbon-based condensed matter physics and materials science.

\section{Acknowledgements}
This work was supported by the National Science Foundation through Grant Nos.~CHE-0809020 and CMS-060950, the Department of Energy BES Program through Grant No.~DEFG02-06ER46308, the Air Force Research Laboratories under contract number FA8650-05-D-5807, the Los Alamos National Laboratory LDRD Program, the Robert A. Welch Foundation through Grant Nos.~C-1509 and C-0807, the National Aeronautics and Space Administration (NNJ05HI05C), and the World Class University Program at Sungkyunkwan University (R31-2008-000-10029-0). This work was performed in part at the Center for Integrated Nanotechnologies, a US Department of Energy, Office of Basic Energy Sciences user facility.  We thank our collaborators who contributed to the work presented in this feature article: B.~Y.~Lu, W.~D.~Rice, S.~Ghosh, R.~B.~Weisman, P.~Nikolaev, S.~Arepalli, C.~G.~Densmore, A.~Jagota, D.~Roxbury, C.~Y.~Khripin and J.~A.~Fagan.  We would also like to thank K.~Yanagi, C.~Kittrell, W.~Adams, N.~Alvarez, C.~Pint, B.~Dan, K.~Sato, R.~Saito, and A.~Imambekov for useful and stimulating discussions.

\end{document}